%% file: 00_main.tex
\documentclass[acmsmall]{acmart}

\AtBeginDocument{%
  }

\newcommand{\para}[1]{\vspace{2pt}\noindent\textbf{#1.~}}
\newcommand{\ignore}[1]{}
\newcommand{\system}{\sloppy{SmartAxe\@}}
\newcommand{\vuln}{\sloppy{CCV\@}}
\newcommand{\bugname}{{\vuln}}

\newcommand{\publicUrl}{\url{https://github.com/InPlusLab/FSE24-SmartAxe}}

\definecolor{verylightgray}{rgb}{.97,.97,.97}

\usepackage{listings, xcolor}
\usepackage{pifont}
\usepackage{ulem}
\usepackage{array,framed}
\usepackage{setspace}
\usepackage{url}

\usepackage{enumerate}
\usepackage{xspace}
\usepackage{multirow}
\usepackage{microtype}

\lstdefinelanguage{Solidity}{
  keywords=[1]{anonymous, assembly, assert, balance, break, call, callcode, case, catch, class, constant, continue, constructor, contract, debugger, default, delegatecall, delete, do, else, emit, event, experimental, export, external, false, finally, for, function, gas, if, implements, import, in, indexed, instanceof, interface, internal, is, length, library, log0, log1, log2, log3, log4, memory, modifier, new, payable, pragma, private, protected, public, pure, push, require, return, returns, revert, selfdestruct, send, solidity, storage, struct, suicide, super, switch, then, this, throw, transfer, true, try, typeof, using, value, view, while, with, addmod, ecrecover, keccak256, mulmod, ripemd160, sha256, sha3}, 
  keywordstyle=[1]\color{blue}\bfseries,
  keywords=[2]{address, bool, byte, bytes, bytes1, bytes2, bytes3, bytes4, bytes5, bytes6, bytes7, bytes8, bytes9, bytes10, bytes11, bytes12, bytes13, bytes14, bytes15, bytes16, bytes17, bytes18, bytes19, bytes20, bytes21, bytes22, bytes23, bytes24, bytes25, bytes26, bytes27, bytes28, bytes29, bytes30, bytes31, bytes32, enum, int, int8, int16, int24, int32, int40, int48, int56, int64, int72, int80, int88, int96, int104, int112, int120, int128, int136, int144, int152, int160, int168, int176, int184, int192, int200, int208, int216, int224, int232, int240, int248, int256, mapping, string, uint, uint8, uint16, uint24, uint32, uint40, uint48, uint56, uint64, uint72, uint80, uint88, uint96, uint104, uint112, uint120, uint128, uint136, uint144, uint152, uint160, uint168, uint176, uint184, uint192, uint200, uint208, uint216, uint224, uint232, uint240, uint248, uint256, var, void, ether, finney, szabo, wei, days, hours, minutes, seconds, weeks, years},  
  keywordstyle=[2]\color{teal}\bfseries,
  keywords=[3]{block, blockhash, coinbase, difficulty, gaslimit, number, timestamp, msg, data, gas, sender, sig, value, now, tx, gasprice, origin},  
  keywordstyle=[3]\color{violet}\bfseries,
  identifierstyle=\color{black},
  sensitive=false,
  comment=[l]{//},
  morecomment=[s]{/*}{*/},
  commentstyle=\color{red}\ttfamily,
  stringstyle=\color{red}\ttfamily,
  morestring=[b]',
  morestring=[b]"
}
\setcopyright{rightsretained}
\acmDOI{10.1145/3643738}
\acmYear{2024}
\copyrightyear{2024}
\acmSubmissionID{fse24main-p195-p}
\acmJournal{PACMSE}
\acmVolume{1}
\acmNumber{FSE}
\acmArticle{12}
\acmMonth{7}
\received{2023-09-29}
\received[accepted]{2024-01-23}

\begin{document}


\title{SmartAxe: Detecting Cross-Chain Vulnerabilities in Bridge Smart Contracts via Fine-Grained Static Analysis}

\author{Zeqin Liao}
\orcid{0000-0003-0306-7465}
\affiliation{%
  \institution{Sun Yat-sen University}
  \city{Zhuhai}
  \country{China}
}
\email{liaozq8@mail2.sysu.edu.cn}

\author{Yuhong Nan}
\orcid{0000-0001-9597-9888}
\affiliation{%
  \institution{Sun Yat-sen University}
  \city{Zhuhai}
  \country{China}
}
\email{nanyh@mail.sysu.edu.cn}

\author{Henglong Liang}
\orcid{0009-0008-9570-1067}
\affiliation{%
  \institution{Sun Yat-sen University}
  \city{Guangzhou}
  \country{China}
}
\email{lianghlong@mail2.sysu.edu.cn}

\author{Sicheng Hao}
\orcid{0009-0009-5747-1093}
\affiliation{%
  \institution{Sun Yat-sen University}
  \city{Zhuhai}
  \country{China}
}
\email{haosch@mail2.sysu.edu.cn}

\author{Juan Zhai}
\orcid{0000-0001-5017-8016}
\affiliation{%
  \institution{University of Massachusetts, Amherst}
  \country{USA}
}
\email{juanzhai@umass.edu}

\author{Jiajing Wu}
\orcid{0000-0001-5155-8547}
\affiliation{%
  \institution{Sun Yat-sen University}
  \city{Zhuhai}
  \country{China}
}
\email{wujiajing@mail.sysu.edu.cn}

\author{Zibin Zheng}
\authornote{Corresponding Author}
\orcid{0000-0002-7878-4330}
\affiliation{%
  \institution{Sun Yat-sen University}
  \city{Zhuhai}
  \country{China}
}
\affiliation{%
  \institution{GuangDong Engineering Technology Research Center of Blockchain}
  \city{Zhuhai}
  \country{China}
}
\email{zhzibin@mail.sysu.edu.cn}

\begin{abstract}

With the increasing popularity of blockchain, different blockchain platforms coexist in the ecosystem  (e.g., Ethereum, BNB, EOSIO, etc.), which prompts the high demand for cross-chain communication. Cross-chain bridge is a specific type of decentralized application for asset exchange across different blockchain platforms. Securing the smart contracts of cross-chain bridges is in urgent need, as there are a number of recent security incidents with heavy financial losses caused by vulnerabilities in bridge smart contracts, as we call them Cross-Chain Vulnerabilities (CCVs). 
However, automatically identifying \vuln{s} in smart contracts poses several unique challenges. Particularly, it is non-trivial to (1) identify application-specific access control constraints needed for cross-bridge asset exchange, and (2) identify inconsistent cross-chain semantics between the two sides of the bridge. 



In this paper, we propose \system{}, a new framework to identify vulnerabilities in cross-chain bridge smart contracts. Particularly, to locate vulnerable functions that have access control incompleteness, \system{} models the heterogeneous implementations of access control and finds necessary security checks in smart contracts through probabilistic pattern inference. 
Besides, \system{} constructs cross-chain control-flow graph (xCFG) and data-flow graph (xDFG), which help to find semantic inconsistency during cross-chain data communication. 
To evaluate \system{}, we collect and label a dataset of 88 \vuln{s} from real-attacks cross-chain bridge contracts. Evaluation results show that \system{} achieves a precision of 84.95\% and a recall of 89.77\%. In addition, \system{} successfully identifies 232 new/unknown \vuln{s} from 129 real-world cross-chain bridge applications (i.e., from 1,703 smart contracts). These identified \vuln{s} affect a total amount of digital assets worth 1,885,250 USD.


\end{abstract}

\ccsdesc[500]{Software and its engineering}
\ccsdesc[500]{Software and its engineering~Software creation and management}
\ccsdesc[500]{Software creation and management~Software verification and validation}

\keywords{Smart Contract, Static Analysis, Cross-chain Bridge, Bug Finding}


\maketitle

\input{01_body}

\normalem

\bibliography{reference}


\end{document}

%% file: 01_body.tex
\section{Introduction}
\label{sec:intro}

The rise of blockchain has prompted a wide range of blockchain platforms (e.g., Ethereum~\cite{Ethereum}, BNB~\cite{BNB}) and crypto-assets (e.g., Bitcoin~\cite{Bitcoin}, Non-Fungible Token~\cite{NFT}). Given such a highly diverse and fragmented ecosystem, there is a strong need for data communication across different blockchain platforms (e.g., exchanging Ether to Bitcoin).  


%
Cross-chain bridge is a specific type of application, working as an intermediary for information exchange (e.g., digital assets) across different blockchains. 
For example, Polygon network bridge~\cite{polygonbridge}, the most popular cross-chain bridge holding more than 2 billion USD, allows users to transfer tokens between Polygon and Ethereum blockchains without other un-trusted third parties.
While gaining a market cap of billions of dollars~\cite{chainspot}, cross-chain bridges face emerging security issues. 
Our investigation showed that in the recent two years, cross-chain bridges have suffered more than 29 security incidents.
A large portion of security incidents in cross-chain bridges are caused by vulnerabilities that reside in their smart contracts. For instance, PolyNetwork~\cite{PolyNetwork} was exploited by an access-control vulnerability, leading to a total loss of 600 million USD~\cite{PolyNetworkexploit}.


In this paper, we call vulnerabilities that are specific to cross-chain bridge smart contracts as \textit{Cross-Chain Vulnerability (\vuln{})}. \vuln{s} are unique to the cross-chain scenario (i.e., asset exchange) implemented by smart contracts. For example, while the root cause of a \vuln{} could be a lack of fine-grained access control, such access control enforcement is not necessary in other traditional smart contracts. As another example, a \vuln{} may caused by inconsistent semantics between the two sides of the bridge (see Section~\ref{sec: problemstatement} for more details).




Given the severe impact of \vuln{s}, there has been very limited research focusing on analyzing \vuln{s}, not to mention a systematic detection framework for securing cross-chain bridges. More specifically, prior research~\cite{duan2023Attacks, lee2023SoK} performed studies to understand the key patterns of cross-chain attacks. Based on their observation and findings, a set of mitigation suggestions were proposed to cross-chain developers. Unfortunately, these suggestions can not be directly applied, or enforced by existing cross-chain bridges. The most relevant work in terms of detecting \vuln{} is Xscope~\cite{zhang2022Xscope}, which can identify a subset of \vuln{s} via anomaly detection. However, Xscope requires analyzing collected statistics of on-chain transactions. With such a prerequisite, it can only detect \vuln{s} that have already been exploited by attackers.


\para{Our Work} In this paper, we propose \system{}, a new static analysis framework to detect \vuln{s} for cross-chain bridge smart contracts. To the best of our knowledge, \system{} is the first of its kind to detect \vuln{s} via program analysis at the bytecode level. In this capability, \system{} enables automatic security vetting for a variety of cross-chain bridge applications before their deployment, and hence, improves their security and amend potential risks that may cause severe damage (e.g., financial losses).



The root cause of \vuln{s} lies in two aspects: 1) access control incompleteness and 2) cross-bridge semantic inconsistency. Identifying \vuln{s} in bridge contracts faces the following two unique challenges in terms of static analysis.

\begin{itemize}
    
    \item Firstly, identifying access control incompleteness relies on the precise extraction of access control constraints. Access control constraints consist of appropriate security checks over specific resources (e.g.,  a critical state variable in smart contract). Unfortunately, such security checks are implemented heterogeneously in bridge contracts. Moreover, associating related resources with such security checks introduces significant complexity. 
    
    \item Secondly, identifying cross-bridge semantic inconsistency heavily relies on analyzing the fine-grained contextual information related to cross-chain communication. For example, inspecting the control-flow and data-flow dependency across multiple blockchains. These information are mostly overlooked by prior frameworks~\cite{tsankov2018Securify, liu2021Smart, liao2023SmartState} when detecting smart contract vulnerabilities. 

\end{itemize}

To this end, \system{} integrates two key designs in its static analysis process for \vuln{} detection. Firstly, to identify necessary security checks in cross-chain bridges,
\system{} models the heterogeneous implementations of access control in bridge contracts to a canonical form. In the meantime, \system{} pinpoints cross-chain related resources that need security checks through probabilistic pattern inference (Section~\ref{sec: AC_construction}). In this way, \system{} can effectively find out those vulnerable functions which have access control incompleteness. 
Secondly, to model the context information, \system{} aligns the control flow and data flow between two sides of the bridge, and further constructs the cross-chain control-flow graph (xCFG) and data-flow graph (xDFG) (Section~\ref{sec: interprocedure}). With the constructed graphs, \system{} locates vulnerable functions containing cross-bridge semantic inconsistency during cross-chain data communication. Lastly, \system{} analyzes the accessibility (i.e., entry point) and subversiveness (i.e., affected state variables) for vulnerable functions, and reports the vulnerability trace (Section~\ref{sec: vul_discovery}).

To evaluate the effectiveness of \system{}, we first construct a manually-labeled dataset (as $D_{manual}$) based on 22 public reports discussing \vuln{}s. The dataset is composed of 16 cross-chain bridge applications, with 88 \vuln{s} from 203 smart contracts. Our experiments show that \system{} is effective in \vuln{} detection with a precision of 84.95\% and a recall of 89.77\% over this dataset. The results show that \system{} effectively detect the majority of \vuln{s} that cause real-world damage.

\para{Detecting \vuln{}s in the wild} With the help of \system{}, we perform a large-scale security vetting of 1,703 smart contracts (from 129 real-world cross-chain bridge applications). To the best of our knowledge, this is the most comprehensive collection of cross-chain bridge smart contracts in the wild. Finally, \system{} reported 232 new \vuln{s} which have not been identified by previous research. The total assets affected by these \vuln{s} reached 1,885,250 USD.

In summary, the contributions of this paper are as follows:

\begin{itemize}

\item We highlight the root causes of cross-chain vulnerabilities (Section~\ref{sec: problemstatement}), including  access control incompleteness and cross-bridge semantic inconsistency.

\item We propose \system{}, the first static analysis framework to detect \vuln{s} for
cross-chain bridge smart contracts.

\item We perform an extensive evaluation to show the effectiveness of \system{}. In addition, by performing a large-scale study over 1,703 cross-chain bridge smart contracts in the wild, \system{} identified 232 new \vuln{s} in the real world.

\item We build the first manual-labeled dataset of cross-bridge vulnerabilities, as well as the most comprehensive dataset of cross-chain bridge applications/smart contracts. To benefit future research, we release the artifact of \system{}, as well as the datasets~\footnote{{\publicUrl}}. 

\end{itemize}


\section{Background and Motivation}
\label{sec:background}

\subsection{Smart Contract and Cross-Chain Bridge}

Smart contract is a specific type of program running on the blockchain. 
These programs support various functionalities to implement new business models~\cite{zheng_overview_2020}, such as decentralized finance (DeFi), decentralized gaming (GameFi), and cross-chain bridges.



\begin{figure}[t]
\centering
\includegraphics[width=2.8in]{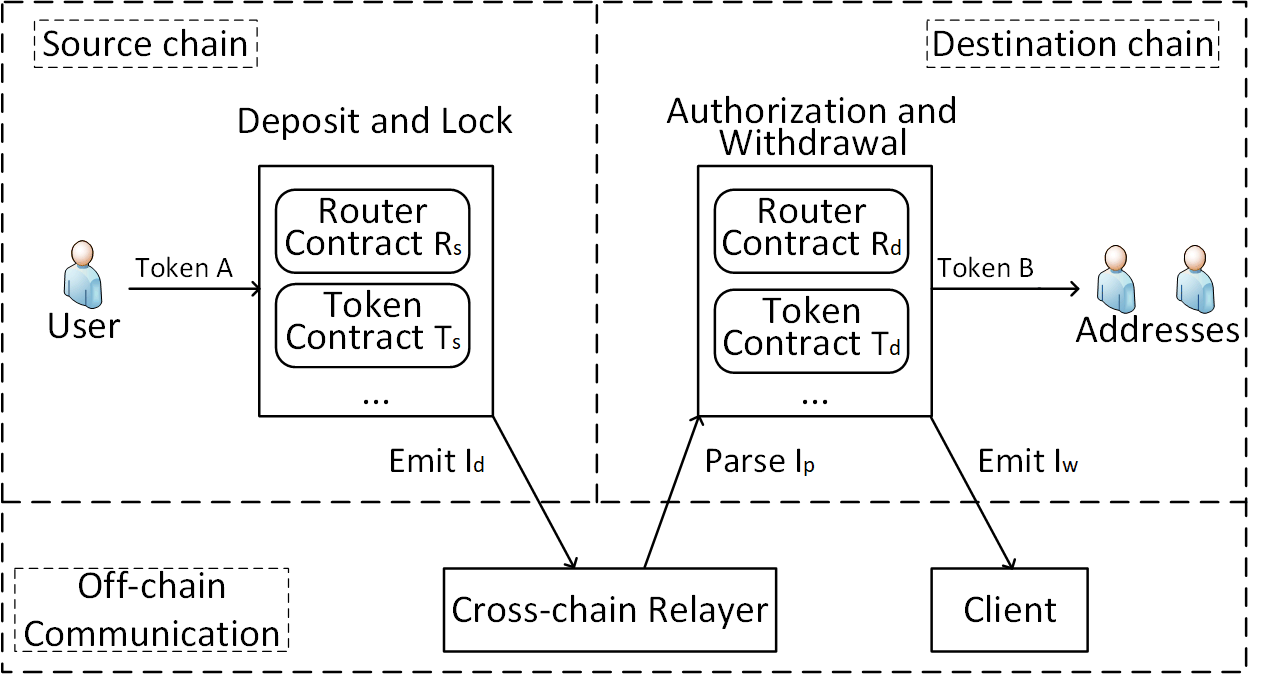}
\caption{Workflow of cross-chain bridge.} 
\label{fig:workflow}
\vspace{-2mm}
\end{figure}

Cross-chain bridge works as an intermediary for asset exchange on different blockchain platforms. Figure~\ref{fig:workflow} shows the key architecture of the cross-chain bridge. A typical cross-chain bridge can be divided into three parts: source chain, cross-chain relayer, and destination chain. The cross-chain bridge deploys smart contracts on the source chain and destination chain. The relayer is designed to support information exchange between the source chain and the destination chain. 
With the cross-chain bridge, users can deposit assets on the source chain and withdraw assets on the destination chain. For example, exchange Token A with Token B in Figure~\ref{fig:workflow}. To detail, this process includes the following three steps:


\begin{enumerate}[(1).]
\item \textbf{Asset deposit on source chain.} When receiving the asset exchange request from the user, the \textit{Router contract} $R_{s}$ of the source chain invokes the \textit{Token contract} $T_{s}$ to lock \textit{Token A}. Then $R_{s}$ emits a deposit event $I_{d}$ as the confirmation of locked assets, which contains the detailed deposit information (e.g., the type and amount). After that, the asset of users would be transferred to the \textit{Router contract} $R_{s}$

\item \textbf{Cross-chain communication via off-chain relayers.} Once a deposit event $I_{d}$ is emitted, the off-chain relayer verifies whether the deposit is valid on the source chain. If the verification gets passed, the relayer transmits the informed information $I_p$ to \textit{Router contract} $R_{d}$ of the destination chain.

\item \textbf{Asset withdrawal on destination chain.} Once $I_p$ is delivered to the destination chain, the \textit{Router contract} $R_{d}$ validates various proofs from $I_p$ for authorization. After the validation, $R_{d}$ emits a withdraw event $I_{w}$ and invokes \textit{Token contract} $T_d$ to withdraw Token B to user-specified addresses on the destination chain.
\end{enumerate}



\subsection{Definition and Problem Statement}
\label{sec: problemstatement}

\begin{figure*}[t]
\centering
\includegraphics[width=5.5in]{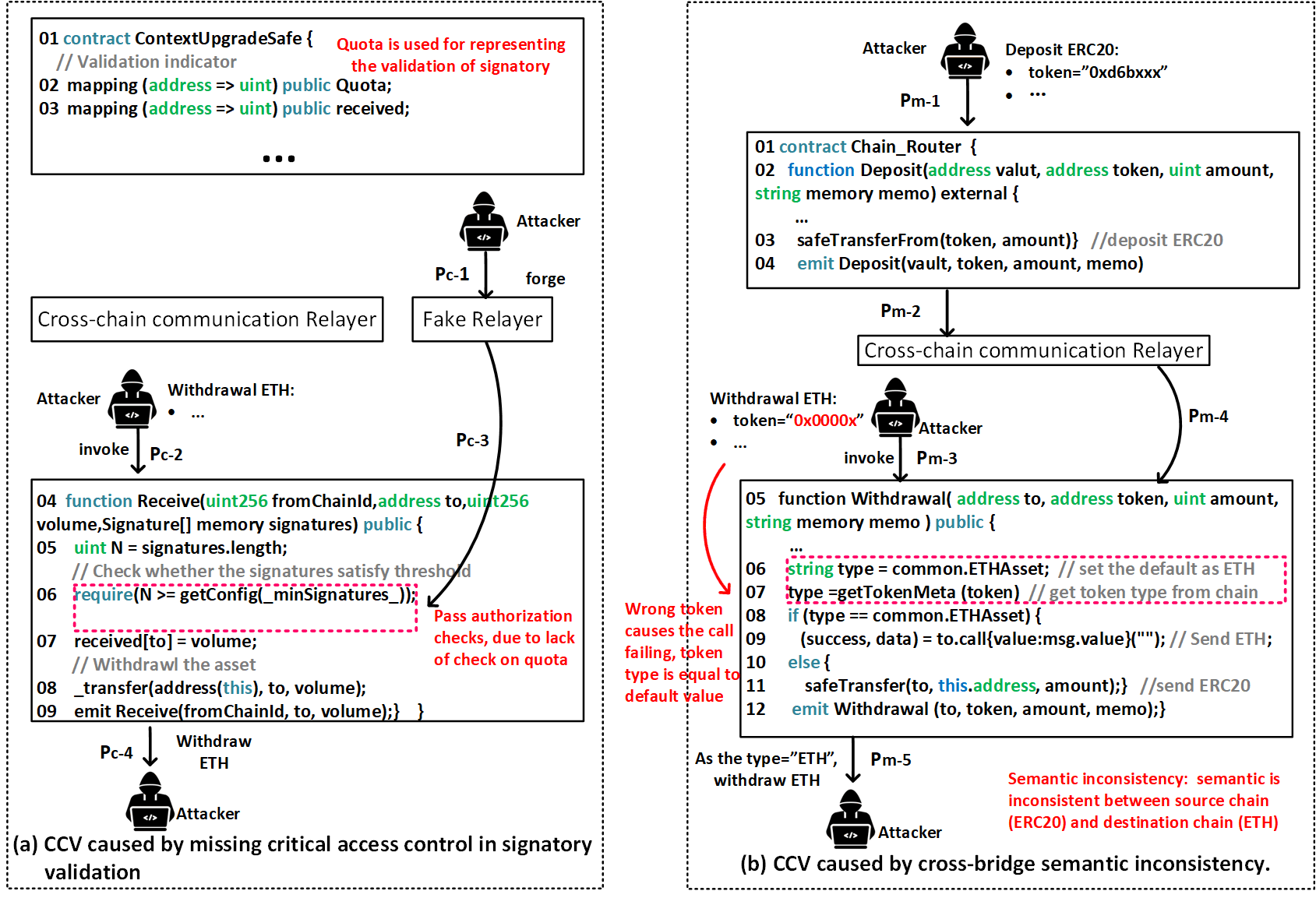}


\caption{Two motivating examples of \vuln{}. }
\label{motivatingexample}
\vspace{-2mm}
\end{figure*}

\para{Cross-Chain Vulnerability} Cross-chain vulnerability (\bugname{}) is a specific type of vulnerability in cross-chain bridge smart contracts. In most cases, cross-chain bridge contract unexpectedly introduces incomplete access control or inconsistent cross-bridge semantics when exchanging assets between the two blockchains. 
%
In the following, we use two motivating examples (Figure~\ref{motivatingexample}) to illustrate \vuln{}. The examples are collected from two real-world bridge contracts that have been exploited by attackers (i.e., ChainSwap~\cite{ChainSwap} and ThORChain~\cite{Thorchain}). We re-organized the original contract code for better illustration.

\begin{itemize}

    \item \para{Access Control Incompleteness} Cross-chain bridge contracts may omit critical security checks, or contain incorrect access control implementations. As shown in Figure~\ref{motivatingexample}(a), variable \textit{Quota} in Contract \textit{ContextUpgradeSafe} represents the validation of signatory (line 2), which should be checked before asset authorization and withdrawal. However, while Function \textit{Receive} is designed for asset authorization and withdrawal, it only checks the signatures without validating the signatory (i.e., checking \textit{Quota}, see the red dotted box (line 6 and 7)). Therefore, the contract introduces an exploitable \vuln{}. As shown in path $P_c$, to attack this cross-chain bridge, the attacker can forge fake relayers (i.e., signatory) for signatures ($P_{c}$-1) and bypass authorization check due to the lack of signatory validation ($P_{c}$-3), and finally withdraw asset ($P_{c}$-4). In practice, this vulnerability can be avoided by adding statements such as \textit{"require (Quota[signatory] > 0)"} between lines 6 and 7.

    \item \para{Cross-bridge Semantic Inconsistency} Ideally, the program semantics of the source chain and destination chain should be aligned with each other, such as type and amount of the exchanged assets. 
    Figure~\ref{motivatingexample} (b) shows an example of semantic inconsistency caused by incorrect parsing of token type. For contract \textit{Chain\_Router}, function \textit{Deposit} emits a record (line 4) of ``ERC-20'' on the source chain. However, the default token expected in function \textit{Withdrawal} is ``ETH'' (line 6). While the actual token type can be updated by \textit{getTokenMeta(token)}, the smart contract on the destination chain does not carefully handle and inspect exceptions. Particularly, an adversary can intentionally pass a wrong token address (i.e., 0x0000x) here, making the token type incorrectly parsed as the default value (``ETH'') regardless of the actual token type on the source chain. Since ETH is more expensive than ERC-20, the attacker can benefit from exchanging assets with such value differences.
    

\end{itemize}

\para{Existing Work and Limitations}
Despite that \bugname{s} have been extensively exploited, there is limited work on identifying this type of vulnerability in advance and further eliminating such losses. To the best of our knowledge, the most relevant work on identifying cross-chain attack is Xscope~\cite{zhang2022Xscope}. However, Xscope is an anomaly detection tool to analyze cross-chain transactions, and it does not support CCV detection at the smart contract level.

Further, as smart contracts are difficult to support patches after deployment, such a framework cannot avoid the attack and further eliminate economic losses.
Additionally, Duan et al.~\cite{duan2023Attacks} and Lee et.al~\cite{lee2023SoK} conduct the survey on cross-chain attacks and propose advice on designing cross-chain systems, so both studies cannot fundamentally detect \bugname{}. 

\subsection{Scope of Our Work}

\system{} is designed to be a generalized framework for detecting \bugname{s} caused by vulnerable smart contracts in cross-chain bridges. 
SmartAxe targets the upper-level application of bridge (i.e., bridge Dapp), rather than the underlying protocol of bridge (e.g., Inter-blockchain communication (IBC) protocol).
As a program analysis framework, \system{} can cover 20 CCV attacks of 29 cross-chain security attacks,
but does not cover other 9 cross-chain exploits irrelevant to smart contract code, such as private key leakage~\cite{RubicPrivateleaking}, DNS hijacking~\cite{CelerNetworkDNSattack} and trusted root leakage~\cite{NOMADRootAttack}.
Given a set of smart contracts in the cross-bridge application, \system{} performs static analysis on the bytecode, and further reports whether the smart contract contains \bugname{}s. Since \system{} conducts analysis on the level of bytecode instead of source code, it is applicable for a number of security vetting scenarios such as large-scale third-party auditing. 



\section{Design of \system{}}
\label{sec:overview}








\begin{figure}[t]
\begin{lstlisting}
contract Radar {
  function bridgeTokens(uint256 amount,bytes32 destChain,address destAddress) external {
    require(balanceOf(msg.sender) >= _amount); // check the balance for ensuring deposit sucess 
    ...
    safeTransferFrom(msg.sender, address(this), amount);
    emit Deposit(amount, destChain, destAddress);}}

contract Polkabridge {
  mapping(address => uint) public balanceOf;
  function mint(address to) external {
    ...
    require(liquidity > 0); // check the liquidity for ensuring deposit sucess
    balanceOf[to] = balanceOf[to] + liquidity;
    emit Deposit(address(0), to, liquidity);}}
\end{lstlisting}
\vspace{-1.8mm}
\caption{An example of the diversity of security checks for deposit success (i.e., lines 3 and 12), which is caused by different implementations and non-standard check, and cannot be addressed by Ethainter~\cite{brent2020Ethainter}, SPCon~\cite{liu2022Findinga}, AChecker~\cite{ghaleb2023AChecker}.}
\label{diversityexample}
\vspace{-3mm}
\end{figure}

\subsection{Challenges and Solutions}
\label{sec: challenges}


With the increasing complexity of cross-chain smart contracts, identifying \bugname{s} is by no means trivial. While prior works conducted analysis on access control vulnerability for smart contracts (e.g., Ethainter~\cite{brent2020Ethainter}, SPCon~\cite{liu2022Findinga}, AChecker~\cite{ghaleb2023AChecker}), these works never consider the security assumption of cross-bridge scenario, therefore, they can hardly detect \bugname{s}. 
Below we list the challenges encountered by \system{} as well as the corresponding solutions. 


\para{C1: Extracting access control constraints}
A typical access control constraint commonly consists of security checks over specific resources in smart contracts. Specifically, security checks can be conditional or compared statements (e.g., \texttt{require}, \texttt{assert}, and \texttt{if}), the resource can be the important operation statement (e.g., read and write on state variables, method invocation). From this perspective, the challenge of extracting access control constraints can be divided into two aspects. The first aspect is the diversity of security checks which is caused by different implementations and non-standard checks. For instance, Figure~\ref{diversityexample} shows the diverse implementation of security checks between different cross-chain bridges. As can be seen, to check the deposit success, contract \textit{Radar} compares the balance of the user with the deposit amount (line 3), contract \textit{Polkabridge} compares the liquidity of the bridge with the threshold (line 12). However, prior research (i.e., Ethainter~\cite{brent2020Ethainter}, SPCon~\cite{liu2022Findinga}, AChecker~\cite{ghaleb2023AChecker}) cannot address this diversity, as their detection patterns are incomplete without considering the semantic of the bridge.

The second aspect is the inherent complexity in linking resources to security checks. We take Figure~\ref{complexityexample} as an example. For the resource of line 12, prior works~\cite{brent2020Ethainter, liu2022Findinga, ghaleb2023AChecker} link this resource to the security checks on which the resource has control flow dependency, i.e., only check of line 5 without check of line 8, which causes the mistake. Actually, by analyzing the semantics on them, we can find that the resource of line 14 transfers the authorized withdrawal, the check of line 5 checks the authorizing signatures, and the check of line 7 checks the authorizing signatory, so all of the checks of lines 5 and 7 should be linked with resource 12. Alternatively, by analyzing the data dependency of lines 8, 9, and 11, we can also determine check of line 7 should be linked with resource 12. However, automatically identifying these is by no means trivial, as it requires analyzing the complex patterns of semantic and data dependency.

To overcome this challenge, for identifying the security checks, we review the documentation and program code of the top 100 cross-chain bridges, and model the access control of cross-chain bridge and normalizes diverse checks to a canonical form (see details via Table~\ref{table:AC_model} and Section~\ref{sec: AC_construction}).
To link the resources to security checks, \system{} utilizes a probabilistic pattern inference method (see detail via Table~\ref{table:pattern} and Section~\ref{sec: AC_construction}). Specifically, \system{} utilizes a set of predefined patterns that consider the dependency of control flow, data flow, and semantics between the resource and security check to determine the association relationship between them. 

Further, based on the extracted access control constraint, \system{} identifies the incompleteness of access control for the bridge contract.

\begin{figure}[t]
\begin{lstlisting}
contract BaseBridge {  
  mapping (address => uint) public authorization;
  function withdrawal(uint256 SrcChain,address to,uint256 amount,Signature[] memory signatures) public {
    ...
    if(signatures.length > _minSignatures); {  //Security Check 1, check the number of signatures
      for(uint i=0; i<signatures.length; i++) {
        address _signatory = getSignatory(signatures[i]);
        require(_signatory == signatures[i].signatory);  //Security Check 2, check the validation signatory 
        authorization[to]=authorization[to]+1; 
        emit Authorize(SrcChain, to, amount, signatory);}}
    require(authorization[to] == signatures.length);  
    _transfer(address(this), to, volume);  // Resource 1, withdrawal the asset
    emit Receive(fromChainId, to, volume);}}

\end{lstlisting}
\vspace{-1.2mm}
\caption{An example to show the complexity in linking resources with security checks.}
\label{complexityexample}
\vspace{-2.0mm}
\end{figure}

\para{C2: Identifying cross-bridge semantic inconsistencies} 
Different from the vulnerability analysis on a single blockchain, identifying \bugname{s} of cross-bridge semantic inconsistency heavily relies on modeling the contextual information (e.g., emitting $I_d$, relayer, informing $I_p$) during cross-chain data transmission. Identifying the contextual information is by no means trivial, as it requires the accurate alignment of control flow and data flow between two sides of the cross-chain bridge, which lack of prior work can support. Further, it is difficult to identify the alignment sites of control flow and data flow, because locating alignment sites requires fine-grained semantic and control flow analysis.



To overcome the second challenge, by the fine-grained semantic and control flow analysis, \system{} identifies two types of functions as the alignment site, i.e., (1) functions that implement deposit and lock and (2) functions that implement authorization and withdrawal.
Then, \system{} aligns the corresponding alignment sites and constructs the cross-chain control flow graph (xCFG). To facilitate the detection for cross-bridge semantic consistency of \bugname{s}, \system{} performs the data flow analysis on the xCFG to construct the cross-chain data-flow graph (xDFG).


Further, based on the constructed xCFG and xDFG, 
\system{} identify the \bugname{} of cross-bridge semantic inconsistency for bridge contracts.



\subsection{Workflow of \system{}}

\begin{figure*}[t]
\centering
\includegraphics[width=5.5in]{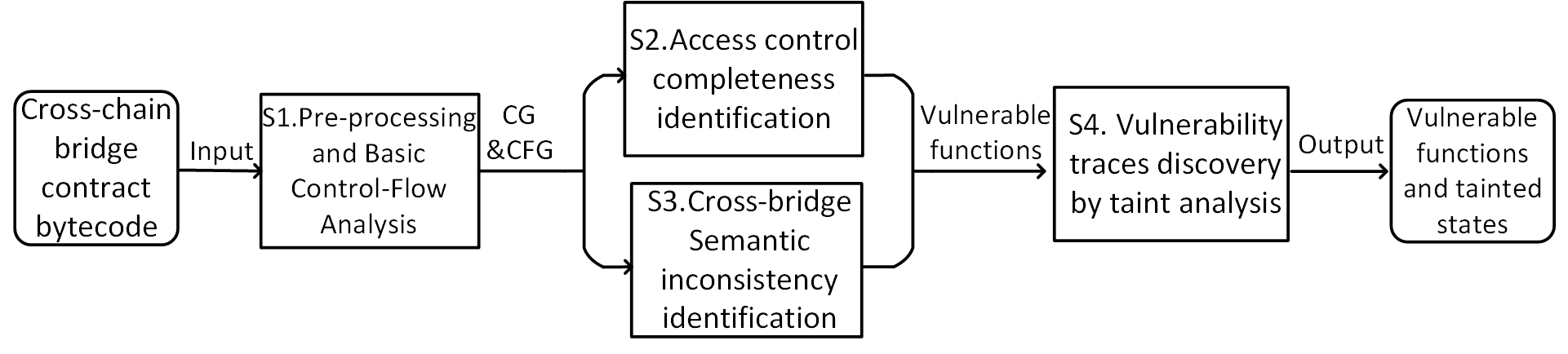}
\caption{The workflow of \system{}.} 
\label{fig:overview}
\vspace{-1mm}
\end{figure*}

\system{} takes the bytecode of cross-chain bridge contracts as its input, and finally reports whether it contains \vuln{s}, as well as the corresponding vulnerable traces. A vulnerable trace contains function calls from the vulnerable function to tainted state variable(s) that can be affected by external call(s). Figure~\ref{fig:overview} demonstrates the workflow of \system{}.






\begin{enumerate} [S1.]
    \item \textbf{Basic control-flow analysis.} As a typical process of static analysis, \system{} separately recovers the control flow of smart contracts on each side of the cross-chain bridge as the pre-processing step, through the existing analyzer (e.g., SmartDagger~\cite{liao2022SmartDagger} in our research).
    
    \item \textbf{Access control completeness identification.} In the second step, \system{} identifies all the access control constraints based on the basic control-flow facts. Particularly, \system{} models the access control of the bridge contracts and normalizes diverse checks to a canonical form. Then, \system{} conducts the probabilistic pattern inference to associate the resources with security checks. Based on the extracted access control constraints, \system{} identifies the vulnerable functions containing incompleteness of access control.
    
    \item \textbf{ Cross-bridge semantic inconsistency identification.} \system{} aligns the control flow of smart contracts between the source chain and destination chain to construct xCFG. Further, \system{} performs data flow analysis to construct xDFG. Based on the constructed graphs, \system{} identifies the vulnerable functions containing semantic inconsistency.
    
    \item \textbf{Vulnerable traces discovery by taint analysis}. Lastly, based on the vulnerable functions reported by S3 and S4, \system{} identifies all the vulnerable traces via taint analysis.
\end{enumerate}

\section{Approach Details}
\label{sec:methodology}

In this section, we demonstrate the details of each step in \system{}. Meanwhile, we utilize a running example (i.e., Figure~\ref{motivatingexample}) to elaborate how \system{} accurately identifies \bugname{s}. 


\subsection{Basic Control-Flow Analysis} 

\system{} separately constructs the basic control flow from the smart contract bytecode on each side of a given cross-chain bridge by utilizing the state-of-the-art static analysis tool SmartDagger. \system{} utilizes SmartDagger which is developed for identifying cross-contract vulnerability, as SmartDagger can construct a more complete control flow for cross-function (contract) invocation compared with other static analysis tools (e.g., Slither~\cite{feist2019Slither} and Mythril~\cite{mythril}). Specifically, for each side of cross-chain bridge, SmartDagger constructs the control flow graph from the bytecode of the bridge contracts.



\subsection{Access Control Incompleteness Identification}
\label{sec: AC_construction}

In this subsection, \system{} extracts the access control constraints by modeling the security checks of access control and associating the resources with security checks. Based on extracted access control constraints, \system{} identifies the incompleteness of access control.

\para{Modeling heterogeneous security checks}
As mentioned earlier, the workflows of cross-chain bridges consist of three steps, 1) asset deposit and lock, 2) cross-chain communication, and 3) asset authorization and withdrawal. The security checks of bridge contracts essentially enforce the permission of users according to such three workflows. However, the implementation for security checks is actually much more diverse in the cross-chain bridge. This is because different cross-chain bridges generally utilize different kinds of security features that are associated with these workflows to realize the security check of access control. Hence, we have to identify, model, and normalize all these access control checks in various forms in order to perform a comparison.


Modeling security checks is based on the fact that, while most of security checks are different in syntactic form, they are semantically equivalent in terms of the protection they provide. \system{} adopt the same definition of equivalence proposed in previous research~\cite{aafer2018AceDroid}. Our definition of equivalence is regarding the protection enabled by the security checks, that is, the kind of malicious behaviors precluded by the security checks. We turn to Figure~\ref{diversityexample} to illustrate this, checks of line 3 and 12 are different from the perspective of syntactic form. However, they are equivalent to each other in terms of the protection they provide, as the check of line 3 ensures sufficient balance for the deposit success and the check of line 12 ensures sufficient liquidity for the deposit success.

To establish requisite access control mechanisms in cross-bridge smart contracts,
we collect the top-100 bridge Dapps from Chainspot~\cite{chainspot}, a website for blockchain bridge aggregation.
Our domain experts meticulously scrutinize these Dapps to discern all security checks embedded within their smart contracts. 
According to the workflows of bridges, our domain experts categorize these security checks accordingly.
For each category, they further stratify them into distinct perspectives based on the targeted protection areas. Subsequently, they provide a comprehensive summary of the corresponding security features and their respective applications associated with each perspective.

Table~\ref{table:AC_model} shows our summarized security check model of access control. \system{} divides the security checks of access control into three categories: (1) category that is specific to asset deposit and lock, (2) category that is specific to cross-chain router, and (3) category that is specific to asset authorization and withdrawal (i.e., the first column). Therefore, we propose to model the security checks of bridge contracts as follows.


\vspace{-0.1in}
\begin{equation}
\small
\begin{split}
    BridgeCheck := &[Deposit and Lock, Crosschain Router, Authorizationandwithdrawal]
\end{split}
\end{equation}

\input{tables/AC_model}

\textbf{Category-1 - asset deposit and lock}: This category includes a success check for the deposit and a validation check for arguments passed by users (i.e., the first two entries of the second column). Specifically, a success check for the deposit is used to confirm the deposit is transferred to the cross-chain bridge and thus prevent fake deposits. The validation check for arguments passed by users is used to prevent users from passing into malicious arguments that cause malicious modification on the contract state. As shown in the first three entries of the third column, security checks of each perspective can adopt various forms, such as comparing the balance of the bridge before and after the deposit, comparing the user balance with the deposit amount, or comparing the liquidity of the bridge with the deposit threshold. Despite the three types of security checks being different in syntactic form, they are semantically equivalent in terms of confirming the success of the deposit, so there is a disjunction relationship between them. In addition, in the fourth and fifth entries of the third column, validation checks for arguments can be implemented by checking the arguments of public function and message invocation (e.g., \textit{msg.sender}, \textit{msg.value}). Similarly, these two types of security checks are semantically equivalent, so there is a disjunction relationship between them. Further, we formulate the above-illustrated understanding as follows.


\begin{equation}
\vspace{-0.1in}
\small
\begin{split}
     Deposit and Lock := &[DepositSuccess, Argument]
\end{split}
\end{equation}

\vspace{-0.15in}
\begin{equation}
\vspace{-0.1in}
\small
\begin{split}
     DepositSuccess := &check(BridgeBalance)\wedge check(UserBalance) \wedge check(AssetOwnership)
\end{split}
\end{equation}

\vspace{-0.15in}
\begin{equation}
\small
\begin{split}
     Argument := &check(FunctionArgument)\wedge check(MessageArgument)
\end{split}
\end{equation}

\textbf{Category-2 - cross-chain routers}: This category mainly checks the correctness of cross-chain routers, as shown in the third entry of the second column. A correctness check for the cross-chain router is used to prevent unexpected logic or errors in the cross-chain data transmission. This type of check can be implemented by (1) checking the supports of the bridge (e.g., Token ID, chain ID) and (2) checking the error of external invocation (e.g. 0 address can cause the external invocation to fail). Obviously, these two types of checks are not semantically equivalent to each other, so \system{} adopts a conjunction relationship between them. We formulate understandings as follows.


\begin{equation}
\vspace{-0.1in}
\small
\begin{split}
     CrosschainRouter := &[Correctness]
\end{split}
\end{equation}

\vspace{-0.15in}
\begin{equation}
\small
\begin{split}
     Correctness := &check(Support) \vee check(ExternalAddress)
\end{split}
\end{equation}

\textbf{Category-3 - asset authorization and withdraw}: This category includes a validation check for authorization, the repetitiveness check, and the correctness check for withdrawal (as shown in the fourth, fifth, and sixth entries of the second column). Specifically, the validation check for authorization is used to ensure that the cross-chain transaction has been signed and proved by relayers. To this end, such validation checks are implemented by checking the correctness and timeout for the signature or signatory. Due to the inequivalence between them, we utilize a conjunction operation to measure them.
The check for repetitive withdrawal is used to avoid users receiving tokens from cross-chain bridge repetitively. For this purpose, this type of check is implemented by inspecting the specific lists (e.g. \textit{mapping variable}) that are used to record the withdrawal.
Further, a correctness check for withdrawal is used to ensure that the transfer targets are valid. This type of security check can be implemented by inspecting the receiver address.   


\begin{equation}
\vspace{-0.1in}
\small
\begin{split}
     Authorizationandwithdrawal := &[AvoidanceofRepetition, AuthorizationVerification,\\& withdrawalCorrectness]
\end{split}
\end{equation}

\vspace{-0.06in}
\begin{equation}
\vspace{-0.1in}
\small
\begin{split}
     AvoidanceofRepetition := &check(RecordedList)
\end{split}
\end{equation}

\vspace{-0.15in}
\begin{equation}
\vspace{-0.1in}
\small
\begin{split}
     AuthorizationVerification := &check(Signature)\vee  check(Timeout)
\end{split}
\end{equation}

\vspace{-0.15in}
\begin{equation}
\small
\begin{split}
     withdrawalCorrectness := &check(RecevierAddress)
\end{split}
\end{equation}



Further, by comparing the control flow with our summarized security check model (Table~\ref{table:AC_model}), \system{} extracts all the security checks of access control for the bridge contracts.


\para{Associating resources with security checks}
After the extraction for security checks, \system{} identifies the resources of access control constraints, and associates the resources with the corresponding security checks through the probabilistic pattern inference method.


\system{} considers four types of resources for access control constraints: (1) FieldAccess, denoted as $f$, (2) Internal method, represented as $m$, (3) application binary interface (ABI), denoted as $a$, and (4) event emitting statement, denoted as $e$. Further, $f$ represents the statements that read or write on state variables (i.e., global variables of smart contracts). $m$ represents the statements that invoke internal methods (e.g. private function) in cross-chain bridge contracts. $a$ refers to the interface of external calls (e.g. public function) in contracts. $e$ refers to the event emitting statements that record the cross-chain data transmission (e.g., deposit record $I_d$). For these four types of resources, the first three types of resources have been well discussed in the existing works~\cite{el-rewini2022Poirot}. Unlike these studies, \system{} introduces the dedicated type of resource (i.e., event emitting statement) for cross-chain bridge scenario, as the most important information of cross-chain transaction is recorded by such event emitting statement.


Before associating the resources with security checks, \system{} first collects basic facts. Specifically, given the resources in bridge contracts $R=\left\{r_1, r_2,..., r_{n1}\right\}$, for each resource $r_i$, \system{} utilizes path-sensitive analysis to compute all reachable paths $P=\left\{p_1, p_2,..., p_{n2}\right\}$. Then, for each reachable path $p_j$, \system{} finds out the security checks $C=\left\{c_1, c_2,..., c_{n3}\right\}$  of access control constraints along the path. Hence, An association between resource $r_i$ and security check $c_k$ on path $p_j$ can be denoted as
$Association(c_k,p_j,r_i)$.

\input{tables/pattern}

Then, \system{} introduces the prior probability to represent the confidence level of the association $Association(c_k,p_j,r_i)$. 
A prior probability is a value between 0 and 1 representing our degree of belief in the association $Association(c_k,p_j,r_i)$.
Furthermore, \system{} determines the prior probability by analyzing access control properties, e.g., control flow, data flow, and semantics between resources and access control checks.
Inspired by prior work~\cite{el-rewini2022Poirot}, we summarize the dedicated association patterns between resources and security checks for cross-chain bridge contracts, as well as their corresponding prior probability, as illustrated in Table~\ref{table:pattern}.


\para{Detecting access control incompleteness} Given the extracted access control constraints of a certain bridge, \system{} identifies the bridge contracts containing \bugname{s} of access control incompleteness and outputs related vulnerable functions, if one of the following aspects is detected.

\begin{enumerate}[(1).]
\item \textbf{Access control omission.} \system{} detects access control omission by comparing the extracted access control constraints with the security check model (Table~\ref{table:AC_model}). If \system{} discovers the omission on the security checks of a certain category (or perspective), \system{} reports the corresponding vulnerable functions.

\item \textbf{Access control violation paths.} 
Similar to prior works~\cite{shao2016Kratos} which are designed for identifying inconsistency access control policies of Android applications,
this aspect is used for identifying access control violation paths that allow users without sufficient permission to access sensitive resources.
For the entry points of bridge contracts, \system{} compares their sub-control-flow graphs pairwise to identify possible paths that can reach the same sensitive resources but enforce different security checks (e.g., one with checks and the other without). \system{} outputs all the vulnerable functions on the access control violation paths.

\end{enumerate}



\subsection{Cross-Bridge Semantic Inconsistency Identification}
\label{sec: interprocedure}

In this subsection, \system{} constructs the xCFG and xDFG by aligning and connecting the single-chain control flow graphs. Further, based on the constructed graph, \system{} identifies the \bugname{} of semantic inconsistency.

\para{Graph Construction}
The xCFG constructed by \system{} can be denoted as $G_c=(N_c, E_c, X_e)$. Specifically, \system{} encodes the following information: (1) the nodes of xCFG is a set of basic block nodes that represent the operations of the program, a relayer node that represents cross-chain data transmission, and a client node that represents the client of the cross-chain bridge. Here, $N_b$ denotes the basic block nodes, $N_r$ denotes the relayer node, and $N_l$ denotes the client node. Therefore, we have $N_c :=\left\{N_b \cup N_r \cup N_l \right\}$; (2) the edges of xCFG are composed of control flow edges $E_f$, emitting edges $E_e$ and informing edge $E_i$. Here, $E_e$ denotes the information flows that the relayer and client watch the emitted events, i.e., the relayer watches the deposit event emitted on the source chain or the client watches the withdrawal event on the destination chain. And $E_i$ denotes the information flows that the relayers inform the contracts of the destination chain to perform authorization and withdrawal. Similarly, we also have $E_c := \left\{ E_f \cup E_e \cup E_i \right\}$; (3) $X_e(E_c) \rightarrow \left\{CF, Emitting, Informing \right\}$ is a labeling function that maps an edge to one of the three types. 



In addition, \system{} constructs the xDFG by performing the data flow analysis on the xCFG. Hence, in terms of data structure, xDFG constructed by \system{} is similar to the traditional data flow graph. The xDFG can be denoted as $G_d=(N_d, E_d)$. Here, $N_d$ denote the different program operation of bridge contracts, and $E_d$ refers to the data dependencies between program operations.


To construct the above-illustrated graphs, \system{} utilizes two key steps for cross-chain inter-procedure analysis.

\system{} constructs xCFG by adding emitting edges and informing edges to single-chain control flow graphs. Specifically, when adding emitting edges, to represent that the relayer watches the deposit event emitted on the source chain, \system{} searches for the emitting statement of deposit and lock event on the source-chain control flow graph as the source of emitting edge, makes the relayer nodes as the target of emitting edge, and connect them with the directed edge. Similarly, to represent that the client watches the withdrawal event on the destination chain, \system{} connects another emitting edge. For such an edge, the source is the emitting statement of authorization and withdrawal event on the destination-chain control flow graph, and the target is the client nodes. When adding the informing edge, to represent that the relayers inform the contracts of the destination chain for authorization and withdrawal, \system{} makes the relayer node as the source, further searches for the authorization statement as the target of the informing edge, and connects them together with the directed edge.


Then, \system{} constructs the xDFG through our proposed dedicated data flow analysis. Similar to traditional data flow analysis,  \system{} performs the forward data flow analysis for control flow edges. Unlike the traditional data flow analysis, for emitting edges, only the arguments of the event can propagate through the emitting edges forward, as only these arguments are recorded in cross-chain data that transmits to the relayer. 
In terms of informing edges, only the arguments invoked by the authorization method can propagate through the informing edge forward. 

\para{Cross-bridge Semantic Inconsistency Detection} Given the constructed xCFG and xDFG of a
certain bridge, \system{} identifies the bridge contracts containing \bugname{s} of semantic inconsistency.
and outputs related vulnerable functions, if one of the following aspects is detected.
\begin{enumerate}[(1).]
    \item \textbf{Semantic granularity check.} The granularity insufficiency of deposit events causes destination-chain contracts cannot distinguish the different types of deposit and enter the same withdrawal logic.
    On xCFG, \system{} detects this aspect by comparing the cross-chain paths pairwise to identify possible paths that converge to the same withdrawal logic on the destination chain but enforce different deposit logic on the source chain. \system{} outputs all the vulnerable functions on the paths containing insufficiently-grained deposit events.
    \item \textbf{Semantic integrity check.} The feature of parse error is that the amount or type of withdrawal depends on the withdrawal function itself rather than the source-chain deposit. 
    On xDFG, \system{} detects such an aspect by identifying whether the state variables of withdrawal have the data-flow dependency on state variables of deposit. If \system{} finds the lack of data-flow dependency between them, \system{} reports corresponding functions as vulnerable.
\end{enumerate}





\subsection{Vulnerability Trace Discovery}
\label{sec: vul_discovery}

In this subsection, \system{} model the vulnerable functions of access control incompleteness and cross-bridge semantic consistency as \bugname{} indicators, and analyze the accessibility (i.e., entry trace) and subversiveness (i.e., affected state variables) for \bugname{} indicators. Finally, \system{} reports vulnerability traces which contain function calls from the vulnerable function to tainted state variable(s).

While vulnerable functions are well identified by Section~\ref{sec: AC_construction} and~\ref{sec: interprocedure}, \system{} still require satisfying the following condition for locating \bugname{s}.

\textbf{Condition-1}: Finding the entry trace of external call. With the \bugname{} indicator, \system{} identifies the entry traces for each \bugname{} indicator. This process is modeled as a process that \system{} utilizes taint propagation to detect whether the \bugname{} indicators can be tainted by external attackers. Specifically, \system{} makes the taint propagate from the entry point (e.g., public function) of the cross-chain bridge contracts, and further detects whether the taint can reach the \bugname{} indicators.

\textbf{Condition-2}: Finding the state variables affected by \bugname{s}. After finding the entry trace of external attackers, \system{} continues to conduct the forward propagation on xDFG, and identify the state variables affected by the \bugname{}. The reason for this step is to find out the state variables that are affected by the subverted flows of the \bugname{}. Lastly, \system{} reports the tainted functions and state variables as various vulnerable traces that reveal how the attackers can leverage the \bugname{}.






\begin{figure*}[t]
\centering
\includegraphics[width=5in]{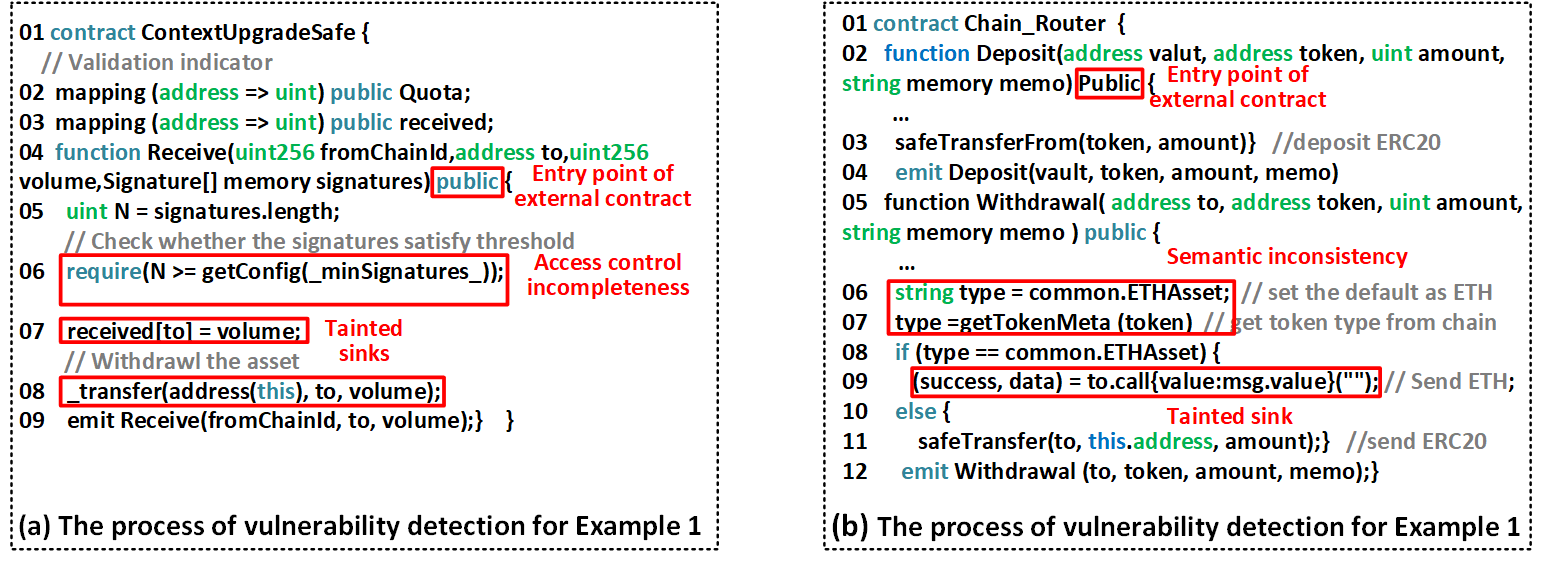}


\caption{The process of vulnerability detection for motivating examples in Figure~\ref{motivatingexample}.}
\label{detection}
\vspace{-2mm}
\end{figure*}

\system{} performs taint propagation via the method proposed in SmartState~\cite{liao2023SmartState}. The taint sources can be divided into two types: the parameters passed by contract callers and the parameters of public functions. The taint sinks of \system{} consist of either those external calls or state variables of the smart contracts
(including the \bugname{} indicators and client node). More detailed information regarding the taint sources and sinks is summarized in table~\ref{table.taint}.

Again, we take the motivating example of Figure~\ref{motivatingexample} as an instance to show the \bugname{} discovery, and the process is shown in Figure~\ref{detection}. For the contract in the left example, \system{} identifies the function \textit{Receive} contains access control incompleteness as the \bugname{} indicator, because \textit{Receive} omits the validation check for signatory (line 6 and 7). Then, \system{} searches for the Condition-2, and identifies that function \textit{Receive} is the entry point of the execution paths, as \textit{Receive} is a public function that can be accessed by an external attacker. Lastly, \system{} investigates the Condition-3 by screening for state variables affected by the vulnerability through taint analysis. As a result, \system{} reports the vulnerability in contract \textit{ContextUPgradeSafe} as follows, $ Receive \rightarrow \_transfer \rightarrow \left\{ received, balance \right\} $. 

Similarly, \system{} identifies vulnerability in contract \textit{Chain\_Router} of right example as follows, $ Deposit \rightarrow Withdrawal \rightarrow \left\{ ETH balance \right\} $

\input{tables/taint}

    

\section{Evaluation}
\label{sec:eval}

In this section, we first introduce the evaluation setup and two datasets used for evaluation (i.e., manually-labeled \bugname{} dataset and large-scale dataset). Then, we show the effectiveness of \system{} and its individual components by evaluating \system{} in terms of precision and recall on the manually-labeled \bugname{} dataset. Lastly, we evaluate the real-world performance of \system{} by conducting the analysis on the large-scale dataset, and identified new \bugname{}s in the wild.


\subsection{Implementation and Evaluation Setup}

We implement \system{} with around 3200 Line-of-Code in Python 3.8.10. Then, we conduct all the evaluation experiments for \system{} on a Ubuntu 20.04 server, which is equipped with the Intel i9- 10980XE CPU (3.0GHz), RTX3090 GPU, and 250 GB RAM.


\para{Dataset and Ground-truth Establishment}
We collect the following datasets for the evaluation experiment.

\textit{Manually-labeled \bugname{} dataset ($D_{manual}$).} We establish the ground truth for evaluating the effectiveness of \system{} in this dataset. Specifically, we collect 16 vulnerable cross-chain bridge applications (i.e., a total of 203 smart contracts) according to 20 real \bugname{} attacks reported by public news and reports. 
Then, we annotate CCVs in the bridge contract by reviewing such attack reports. Specifically, each CCV is determined by two essential parts, (1) a vulnerable smart contract and (2) one or multiple vulnerable traces trigger the vulnerability. Here, a vulnerable trace contains function calls from the vulnerable function to tainted state variable(s) that can be affected by external call(s). In this way, we annotate a total of 88 CCV traces for 16 vulnerable bridge applications.
To avoid bias, we invite three domain experts for the manual annotation, each expert separately performs the annotation. Only the vulnerability agreed by all three experts is confirmed as a valid \bugname{}.  
To the best of our knowledge, this is the most comprehensive collection of \bugname{}s from public sources. 

\textit{Large-scale dataset ($D_{large}$).} We collect the large-scale dataset to show the effectiveness of \system{} in identifying \bugname{s} in the wild. The large-scale dataset contains 129 cross-chain bridge applications (i.e., a total of 1703 smart contracts). Specifically, we searched for the cross-chain bridge project in the community and Internet exhaustively, and finally looked up a total of 148 cross-chain bridge applications. Further, we manually confirm these applications, and find out both the source code and bytecode for 129 cross-chain bridge applications among them (i.e.,  the coverage rate is over 87\%). Finally, we comb the source code and bytecode of these 129 cross-chain bridge applications and obtain the large-scale dataset. 
Our collection actually covers the mainstream of cross-chain bridges. For example, 86 of 129 cross-chain bridges are contained in the top 100 cross-chain bridges ranked by the Chainspot~\cite{chainspot}. To the best of our knowledge, this is the most comprehensive collection of cross-chain bridge contracts from public sources.



\para{Evaluation Metrics}
We lay out the following research questions (RQs) for the evaluation experiments.

\begin{itemize}
    \item RQ1. How does \system{} perform in detecting \vuln{s}?
    \item RQ2. How effective is \system{} in finding access control incompleteness?
    \item RQ3. How effective is \system{} in finding cross-bridge semantic inconsistencies?
    \item RQ4. Can \system{} detect \bugname{s} from real-world cross-chain bridge applications?
\end{itemize}


\input{tables/evaluation_overall}

\subsection{Effectiveness of \system{}}

To answer RQ1, we run \system{} over the manually-labeled \bugname{} dataset and evaluate its precision and recall rate. Particularly, we give the same time budget (i.e., 10-mins timeout following prior works~\cite{liao2022SmartDagger,liao2023SmartState}) for every smart contract in the dataset. More specifically, we compute the precision and recall rate by manually comparing the results reported by \system{} with the ground truth of $D_{manual}$ (i.e., 88 \bugname{s} in 203 cross-chain bridges). Note that we do not conduct comparison experiments, as \system{} is the first \bugname{} detection approach for cross-chain bridge contracts.

Table~\ref{table.overall} demonstrates the precision and recall of \system{}. As can be seen, \system{} achieves high precision (i.e., 84.95\%) and recall (i.e., 89.77\%) for \bugname{} detection. To this end, we can conclude that \system{} can identify \vuln{s} for cross-chain bridge contracts effectively.

\para{False Positives and False Negatives} We manually investigate all the false positives and false negatives reported by \system{}. Our investigation result shows that most of the 14 false positives
are caused by the limitation of basic facts produced by SmartDagger (i.e., the basic control-flow analyzer in \system{}). For instance, SmartDagger is insufficiently precise in recovering the type and semantics of state variables for smart contracts, which causes \system{} to report false positives. To overcome these false positives, \system{} can integrate a more advanced analyzer to improve the effectiveness of type and semantic recovery. In terms of false negatives, the reason for them is that 
\system{} ignores a tiny amount of access control that relies on the on-chain query (e.g., checking ownership of on-chain asset). Actually, such problems are unable to be solved through a static analysis approach like ours, because they require real-time on-chain data.




\subsection{Impact of Security Check Modeling and Resource Association}

To answer RQ2, we evaluate the effectiveness of the security check modeling and resource association respectively. 

\para{Effectiveness of Security Check Modeling} As illustrated in Section~\ref{sec: AC_construction}, security check modeling is the important advantage of \system{}. The security check modeling helps for covering the diverse security checks and facilitates the discovery for access control omission effectively, so \system{} can utilize such advantage to identify more vulnerable traces. The security check modeling ensures soundness (i.e., avoiding false negatives) for vulnerability analysis. To evaluate this, we compare the recall rate of \system{} with two state-of-the-art tools (i.e. Ethainter~\cite{brent2020Ethainter}, Achecker~\cite{ghaleb2023AChecker}). Note that we do not compare \system{} with SPCon, as AChecker has been proven to perform better than SPCon in prior work~\cite{ghaleb2023AChecker}. 
We ran \system{}, Ethainter, and AChecker over the manually-labeled dataset $D_{manual}$ to compare their recall rate.

\input{tables/checkextraction}
\input{tables/resource_allocation}

As can be seen in table~\ref{table.securitycheckextraction}, the recall rate of \system{} is much higher compared with the other two tools. To identify why \system{} performs better, we manually inspect all the false negatives for the other two state-of-the-art tools. Particularly, the manual inspection results demonstrate that, most false negatives are caused by the limitation of these state-of-the-art tools, as they cannot overcome the diversity of security checks in cross-chain bridge contracts. In contrast, \system{} can utilize the proposed normalized access control model ( i.e., Table~\ref{table:AC_model}) to avoid these false negatives. To illustrate, we take the motivating example of Figure~\ref{motivatingexample} (a) as an instance again, which has been discussed in Section~\ref{sec: problemstatement} earlier. Noting that Ethainter and AChecker produce false negatives for Figure~\ref{motivatingexample}, but
\system{} can avoid such false negatives by security check modeling.  With the security check model of Table~\ref{table:AC_model}, \system{} identifies that function \textit{Receive} omits the validation check on signatory by comparing function \textit{Receive} with perspective P4 of Table~\ref{table:AC_model}, and reports a \bugname{} of access control incompleteness.


\para{Effectiveness of Resource Allocation} Another advantage of \system{} is that \system{} reduces the false positives by associating resources with the security checks. To evaluate this, we compare the \system{} with \system{} without resource allocation, i.e., \system{} without resource association. To evaluate the effectiveness of resource association. We ran the \system{} without resource allocation and \system{} over manually-labeled dataset $D_{manual}$.

Table~\ref{table.resource_allocation} presents the precision of \system{} and \system{} without resource allocation. Due to unloading the resource allocation method, the precision of the \system{} without resource allocation is only 77.45\%, and the \system{} without resource allocation reports more false positives (i.e., 11 new false positives). To this end, we can summarize that the resource allocation method helps \system{} improve the precision for \bugname{} detection.

Further, we manually inspect each false negative reported by the \system{} without resource allocation. The inspection results demonstrate that 11 of 23 false positives (i.e., 47.83\%) can be eliminated by utilizing the resource allocation method, which are missed by the \system{} without resource allocation. 
To illustrate, we take the motivating example of Figure~\ref{complexityexample} as an instance again, which has been discussed in Section~\ref{sec: challenges} earlier.
\system{} without resource allocation 
produces false positives for Figure~\ref{complexityexample}, 
and report function \textit{withdrawal} omits validation check for signatory, as it cannot identify the association between the security check of line 8 and resource of line 12.
With the probabilistic inference pattern of resource allocation in Table~\ref{table:pattern}, \system{} identifies that security checks of line 5 and 8 should be associated with the resource of line 12, and further eliminate such false positives.




\subsection{Impact of Graph Construction}

As illustrated in Section~\ref{sec: interprocedure}, another advantage possessed by \system{} is our constructed comprehensive xCFG and xDFG, which facilitate
the detection for \bugname{} of semantic inconsistency.

Hence, the constructed graphs help for performing more semantic inconsistency discovery and taint tracking,
so that it can identify more vulnerability traces. Therefore, the effectiveness of graph construction reflects on the recall rate. Similarly, we compare the \system{} with \system{} without xCFG and xDFG construction, to evaluate the effectiveness of graph construction. We ran all of these tools on the manually-labeled dataset ($D_{manual}$) to evaluate their recall.

Table~\ref{table.inter_procedure} shows the comparison results between \system{} and the \system{} without graph construction. Due to ignoring graph construction, the recall rate of the \system{} without graph construction is only 65.91\% and drops rapidly. In contrast, \system{} presents a better performance (i.e., 89.77\%). In conclusion, the constructed xCFG and xDFG help \system{} improve the recall of \vuln{} detection effectively.

\input{tables/inter_procedure}

In addition, we manually investigate all the false negative results for \system{} without graph construction. The manual analysis results show that 23 of 32 false negatives actually can be eliminated through performing end-to-end analysis on xCFG and xDFG, which are missed by the \system{} without graph construction. 



\subsection{Cross-Chain Vulnerability in Real world}

To answer RQ4, we run \system{} over the large-scale dataset ($D_{large}$) to evaluate the real-world performance of \system{}. Our domain experts manually inspected all the reported results via majority voting, and finally confirmed that \system{} detects 232 new \bugname{} from 129 cross-chain bridge applications (i.e., affecting 126 smart contracts). To detail, \system{} outputs 324 warnings. Among them, our manual investigation showed that 278 are true positives and the rest 46 are false positives. 
By inspecting the amount of assets affected by these \bugname{}, we find that these 232 new \bugname{s} affect a total asset of 1,885,250 USD as of our paper submission. Figure~\ref{table.bridge} presents the top 5 cross-chain bridges in terms of assets affected by \bugname{}. For more detail, we
discuss two case studies for illustration. 

\para{Case study 1} at \textit{0x915861959D2feBCCF37795Fd93c6094DdeBf34Bd}. 
This smart contract is from a real-world bridge ranking in the top 40 cross-chain bridges~\cite{chainspot}. However, such a smart contract contains a \bugname{} of access control incompleteness. Specifically, this contract implements the authorization and withdrawal via two functions, i.e., function \textit{saveWithdrawNative} for native tokens of the bridge and function \textit{saveWithdrawAlien} for alien tokens outside the bridge. Unfortunately, both two functions omit the security checks on token type (i.e., native token or alien token). Hence, the deposit record of the native token can unexpectedly pass the authorization of function \textit{saveWithdrawAlien}, and cause the alien token to be withdrawn, 
and it is the same the other way around.
For this case,  by comparing functions (i.e., \textit{saveWithdrawNative} and \textit{saveWithdrawAlien}) with perspective P3 of Table~\ref{table:AC_model}, \system{} effectively identifies them as vulnerable functions and utilizes the taint analysis to determine that 
state variable \textit{balance} is manipulated by the \bugname{}.

\para{Case study 2} at \textit{0x2d6775C1673d4cE55e1f827A0D53e62C43d1F304}. This smart contract is from a real-world bridge, which ranks in the top 30 cross-chain bridges~\cite{chainspot}. Particularly, this contract also involves a \bugname{} of access control incompleteness. Specifically, such a contract implements the deposit of liquidity via function \textit{preFill}. Particularly,  \textit{preFill} receives a calldata argument \textit{\_message} which contains the sender, destination, and amount of the deposit. Unfortunately, \system{} omits the validation check on such an argument. Hence, an attacker can construct the fake \textit{\_message} to obtain a malicious deposit. 
For this case,  by comparing functions (i.e., \textit{preFill}) with perspective P2 of Table~\ref{table:AC_model} , \system{} effectively identifies function \textit{preFill} as vulnerable functions and utilizes the taint analysis to determine that 
state variable \textit{balance} and \textit{liquidityProvider} is manipulated by the \bugname{}.


\input{tables/real-world}



\subsection{Discussion and Limitation}

\system{} shares the following advantages in \vuln{} detection: (1) As demonstrated in evaluation, \system{} is obviously effective in locating \vuln{s} for cross-chain bridge contracts, which lack of prior works can support~\cite{zheng2023survey}; (2) \system{} establishes comprehensive extraction for access control constraints, and cover the diverse and complex access control for cross-chain bridge contract effectively, which overcome the limitation of prior works~\cite{brent2020Ethainter,ghaleb2023AChecker,liu2022Findinga}; (3) \system{} propose unique a graph construction (i.e., xCFG and xDFG) to locate the \vuln{s} of semantic inconsistency effectively, which lack of prior works can support. With the above-illustrated advantages, \system{} can identify \vuln{} precisely and comprehensively. All of the developers, participants, and third-party authorities can utilize \system{} to investigate the security of cross-chain bridge contracts.


While \system{} leverages predefined patterns (Table~\ref{table:AC_model}) and probability values (Table~\ref{table:pattern}), it remains a highly generalizable framework for CCV detection. This is attributed to the following reasons: (1) The predefined patterns are derived from three fundamental workflows of the bridge, namely asset deposit, cross-chain communication, and asset withdrawal. These security checks are essential and universally applicable across all bridges. (2) In terms of comprehensiveness, our investigation covers the top 100 bridges, which represent 67.57\% (100/148) of bridges in existence, thereby affecting over 90\% of all cross-chain transactions. Given the prevalence and significance of these top bridges, the proposed patterns adequately address most CCVs. (3) The probability values are derived from the control-flow and data-flow dependencies inherent in the bridge contract. As all bridges rely on similar dependencies (such as reachability and variable dependency) to facilitate resource protection, the probability values hold true across all bridges.

\para{Responsible disclosure} To prevent real economic damage to the bridge contracts, we anonymize the reported results in this paper and only showcase the Top 5 cross-chain bridges in terms of assets affected by our confirmed CCVs, as illustrated in Table~\ref{table.bridge}.

We are gradually reporting the identified CCVs to bridge developers, a process that is time-consuming and nontrivial. Confirming such CCVs ourselves poses significant challenges: (1) Conducting on-chain testing for vulnerability confirmation may lead to economic loss or damage. (2) Off-chain simulation for confirming CCVs necessitates setting up the entire cross-bridge infrastructure (e.g., relayers and underlying protocols), demanding substantial engineering effort. Presently, all reports are awaiting acknowledgment by the corresponding parties such as project developers. 

\para{Threats to validity} Below we discuss the soundness and completeness of individual components of \system{}. (1) For security check extraction, the completeness of \system{} is threatened by the insufficiently precise basic fact (produced by SmartDagger); (2) For the resource allocation, the soundness of \system{} is influenced and subverted by the uncertainty of the probabilistic method.
(3) The semantic inconsistency identification and vulnerability trace location part of \system{} are sound and complete, because they neither introduce false information nor miss valid information.

Note that \system{} currently lacks support for analyzing non-EVM chains, as its core control-flow analyzer, SmartDagger, is specifically designed for EVM chains. Nevertheless, \system{} offers the flexibility to integrate alternative analyzers tailored for non-EVM chains to extend its capabilities in supporting such networks.
Importantly, the effectiveness of \system{} remains unaffected by this limitation for the following reasons:
(1) The vast majority of bridge Dapps predominantly operate on or support EVM chains, as evidenced by prior research~\cite{lee2023SoK}.
(2) The smart contracts deployed by bridge Dapps across their supported chains, including both EVM and non-EVM chains, exhibit consistent program logic. Therefore, the absence of support for non-EVM chains does not compromise the efficacy of \system{} in detecting vulnerabilities.



\vspace{-1mm}

\section{Related Work}

\para{Smart Contract Vulnerability Detection}
In recent years, many security tools have been proposed to detect vulnerabilities in smart contracts. They can be divided into static analysis tools and dynamic analysis tools. 
For example, static tools include Oyente~\cite{luu2016Making}, MadMax~\cite{grech2018Madmax}, Securify~\cite{tsankov2018Securify},  Zeus~\cite{kalra2018ZEUS}, Clairvoyance~\cite{xue2020Crosscontract}, \textsc{SailFish}~\cite{bose2022SAILFISHa}, eTainter~\cite{ghaleb2022ETainterb}, SmartState~\cite{liao2023SmartState}, etc. Other tools such as ContractFuzzer~\cite{jiang2018ContractFuzzer}, Harvey~\cite{wustholz2020Harvey}, Echidna~\cite{grieco2020Echidna}, sFuzz~\cite{nguyen2020SFuzz}, SMARTIAN~\cite{choi2021SMARTIAN}, and ItyFuzz~\cite{shou2023ItyFuzz} belong to dynamic analysis or testing. As for access control in smart contracts, early research like Mythril~\cite{mythril} only focuses on unprotected Ether withdrawal and specific instructions. While recent works such as Ethainer~\cite{brent2020Ethainter}, SPCon~\cite{liu2022Findinga}, and AChecker~\cite{ghaleb2023AChecker} consider the access control policy or model, they are limited to analyzing contracts of a single address. Therefore, These frameworks cannot identify access control constraints of cross-chain bridges that involve multiple addresses. As all these tools do not consider cross-chain analysis, they are not effective in detecting \vuln{s}. 

\para{Cross-chain Attack Detection}
Our work is also closely related to various cross-chain attack detection mechanisms. Cross-chain systems have become necessary infrastructures for decentralized applications~\cite{lee2023SoK}. Previous work mainly focuses on the security of designing a cross-chain system, such as zkBridge~\cite{xie2022ZkBridge} and CrossLedger~\cite{vishwakarma2023CrossLedger}. In addition, Duan et.al~\cite{duan2023Attacks} and Lee et.al~\cite{lee2023SoK} conduct the survey on cross-chain attacks and propose advice on designing cross-chain systems. These studies discuss the severity of cross-chain attacks and vulnerabilities. 
To the best of our knowledge, the most relevant work on identifying \vuln{} is Xscope~\cite{zhang2022Xscope}. It collects the transactions and status on different blockchains for existing attack detection with a set of security properties and patterns. Nonetheless, as smart contracts are hard to fix after deployment, such a framework cannot find vulnerabilities in contract code before deployment and further eliminate economic losses. Our proposed framework, \system{}, is the first of its kind to detect cross-chain vulnerabilities in smart contracts.

\vspace{-1mm}

\section{Conclusion}
\label{sec:conclusion}

In this paper, we propose a new static analysis framework, \system{}, for locating cross-chain vulnerabilities in cross-chain bridge contracts. 
We evaluate \system{} over a manually-labeled dataset of 16 real attack cross-chain bridge applications (i.e., 203 smart contracts) and a large-scale dataset of 129 real-world cross-chain bridge applications (i.e., 1703 smart contracts). Evaluation results show that \system{} can identify cross-chain vulnerability effectively, with a high precision of 84.95\% and a high recall of 89.77\%. In addition, we find that 232 new \vuln{s} exist in 126 on-chain smart contracts, affecting a total asset of 1,885,250 USD.

%


\begin{acks}
\label{sec:Acknowledgements}
The research was supported by the National Natural Science Foundation of China (62032025), and the Technology Program of Guangzhou, China (No. 202103050004).
\end{acks}

%% file: tables/AC_model.tex
\begin{table*}[t]
\scriptsize
\caption{Security check model of access control for cross-chain bridge contract.}


\begin{tabular}{l|l|l|l}
\hline
Category                                                                                   & Perspective                                                                                            & Security feature                & Example of usage                                                                                               \\ \hline
\multirow{5}{*}{\begin{tabular}[c]{@{}l@{}}C1. Asset deposit \\ and  locking\end{tabular}}         & \multirow{3}{*}{\begin{tabular}[c]{@{}l@{}}P1.Success check for\\  the deposit\end{tabular}}           & Balance of bridge after deposit & Comparison with balance before deposit                                                                         \\ \cline{3-4} 
                                                                                           &                                                                                                        & Balance of user                 & Comparison with the deposit amount                                                                             \\ \cline{3-4} 
                                                                                           &                                                                                                        & Liquility of bridge             & Comparison with deposit threshold                                                                              \\ \cline{2-4} 
                                                                                           & \multirow{2}{*}{\begin{tabular}[c]{@{}l@{}}P2.Validation check \\ for arguments of user\end{tabular}} & Arguments of public function    & Comparison with logic condition                                                                                \\ \cline{3-4} 
                                                                                           &                                                                                                        & Arguments of user message       & Comparison with logic condition                                                                                \\ \hline
\multirow{2}{*}{\begin{tabular}[c]{@{}l@{}}C2.Cross-chain\\  router\end{tabular}}          & \multirow{2}{*}{\begin{tabular}[c]{@{}l@{}}P3.Correctness check\\ for cross-chain router\end{tabular}} & Bridge-supported token/chain & Comparison with ID of destination                                                                              \\ \cline{3-4} 
                                                                                           &                                                                                                        & Address of external invocation  & Comparison with 0 address                                                                                      \\ \hline
\multirow{7}{*}{\begin{tabular}[c]{@{}l@{}}C3. Asset \\ authorization\\ and withdrawal\end{tabular}} & \multirow{2}{*}{\begin{tabular}[c]{@{}l@{}}P4.Validation check \\ for verification\end{tabular}}       & Signature and Signatory         & Comparison with cross-chain message                                                                    \\ \cline{3-4} 
                                                                                           &                                                                                                        & Timeout of signature            & \begin{tabular}[c]{@{}l@{}}Comparison with on-chain time status\\  (e.g., timestamp, blocknumber)\end{tabular} \\ \cline{2-4} 
                                                                                           & \begin{tabular}[c]{@{}l@{}}P5.Check for \\ repetitive withdrawal\end{tabular}                          & List recording the withdrawal   & \begin{tabular}[c]{@{}l@{}}Consultation on the lists (i.e., mapping \\ variables)\end{tabular}                 \\ \cline{2-4} 
                                                                                           & \begin{tabular}[c]{@{}l@{}}P6.Correctness check\\ for releasing\end{tabular}                           & Recevier address                & \begin{tabular}[c]{@{}l@{}}Comparison with user-specified address \\ or 0 address\end{tabular}                 \\ \hline
\end{tabular}

\label{table:AC_model}
\end{table*}

%% file: tables/pattern.tex
\begin{table*}[t]
\scriptsize
\caption{Probabilistic inference pattern for associating resources with security checks.}

\begin{tabular}{lll}
\hline
Pattern & Condition                                 & Probabilistic assignment                                           \\ \hline
P1      & $ControlFlowDependency(c,\{r\})$            & $Association(c,p,r) = true (0.95)$                                   \\
P2      & $ControlFlowDependency(c,R) \vee r\in R $ & $Association(c,p,r) = true (0.60)$                                    \\
P3      & $SameBlock(r_1,r_2)$                          & $Association(c,p_1,r_2) \stackrel{0.60}{\longrightarrow}  Association(c,p_1,r_1) $  \\
P4      & $SemanticCorrelation(r_1,r_2)$                & $Association(c,p_2,r_2) \stackrel{0.70}{\longrightarrow} Association(c,p_1,r_1) $  \\
P5      & $DataFlowDependency(r_1,r_2)$                 & $Association(c,p_2,r_2) \stackrel{0.80}{\longrightarrow} Association(c,p_1,r_1) $ \\ \hline
\end{tabular}

\label{table:pattern}
\end{table*}

%% file: tables/taint.tex
\begin{table}[h]
\scriptsize
\caption{EVM instructions defined as taint sources and taint sinks by \system{}.}

\begin{tabular}{l|l|l}
\hline
                        & Type                                     & EVM Instruction or Keyword or Statement                                                                        \\ \hline
\multirow{2}{*}{Source} & (1) Parameter passed by user & \begin{tabular}[c]{@{}l@{}}CALLDATALOAD, CALLDATACOPY, CALLER, ORIGIN, \\ CALLVALUE, CALLDATASIZE\end{tabular} \\ \cline{2-3} 
                        & (2) Parameter of public function         & Public, External                                                                                               \\ \hline
\multirow{2}{*}{Sink}   & (1) External calls                       & CALL, CALLCODE, STATICCALL, DELEGATECALL                                                                       \\ \cline{2-3} 
                        & (2) State variables                      & SSTORE, BALANCE, ADDRESS, \vuln{} indicators, Client node                                                          \\ \hline
\end{tabular}


\label{table.taint}
\vspace{-2mm}
\end{table}

%% file: tables/evaluation_overall.tex
\begin{table}[bp]
\scriptsize
\caption{Overall effectiveness for \system{} on the Manual-labelled \vuln{} Dataset ($D_{manual}$).}

\begin{tabular}{l|ccc|ccc}
\hline
 \vuln{}                                  & \multicolumn{3}{c|}{Precision}                                               & \multicolumn{3}{c}{Recall}                               \\ \hline
                                        & \multicolumn{1}{r}{TP} & \multicolumn{1}{r}{FP} & \multicolumn{1}{r|}{rate} & TP & \multicolumn{1}{r}{FN} & \multicolumn{1}{r}{rate} \\ \hline
\multicolumn{1}{l|}{Access control incompleteness} & 54                       &10                        & 84.38\%                           & 54 & 5                      & 91.53\%                    \\
Semantic inconsistency                              & 25                       & 4                     & 85.71\%                           & 25 & 4                      & 86.21\%                    \\ \hline
Total                                   & 79                       & 14                       & 84.95\%                           & 79 & 9                     & 89.77\%                    \\ \hline
\end{tabular}

\label{table.overall}
\end{table}

%% file: tables/checkextraction.tex
\begin{table}[htbp]
\scriptsize

\caption{Effectiveness of security check modeling with other SOTA static analyzers over $D_{manual}$.}

\begin{tabular}{l|ccc|ccc|ccc}
\hline
Approach   & \multicolumn{3}{c|}{Ethainter~\cite{brent2020Ethainter}} & \multicolumn{3}{c|}{AChecker~\cite{ghaleb2023AChecker}} & \multicolumn{3}{c}{\system{}} \\ \hline
           & TP     & FN    & Recall    & TP      & FN     & Recall     & TP    & FN    & Recall     \\ \hline
\vuln{}  &6        &82       &6.82\%           &3         &85        &3.41\%            & 79    & 9    & 89.77\%    \\ \hline
\end{tabular}

\label{table.securitycheckextraction}
\end{table}

%% file: tables/resource_allocation.tex
\begin{table}[htbp]
\scriptsize
\caption{Comparison results between \system{} and \system{} without resource allocation on the Manual-labelled \vuln{} Dataset ($D_{manual}$), for evaluating the effectiveness of resource allocation }

\begin{tabular}{l|ccc|ccc}
\hline
Approach   & \multicolumn{3}{c|}{\system{} w/o resource allocation} & \multicolumn{3}{c}{\system{}} \\ \hline
           & TP             & FP            & Precision              & TP    & FP    & Precision      \\ \hline
\vuln{} & 79             & 23            & 77.45\%            & 79    & 14    & 84.95\%    \\ \hline
\end{tabular}

\label{table.resource_allocation}
\end{table}

%% file: tables/inter_procedure.tex
\begin{table}[htbp]
\scriptsize
\caption{Impact of graph construction over $D_{manual}$.}

\begin{tabular}{l|ccc|ccc}
\hline
Approach   & \multicolumn{3}{c|}{\system{} w/o xACG and xDFG} & \multicolumn{3}{c}{\system{}} \\ \hline
           & TP               & FN              & Recall               & TP    & FN    & Recall     \\ \hline
\vuln{} & 56               & 32              & 65.91\%              & 79    & 9     & 89.77\%    \\ \hline
\end{tabular}

\label{table.inter_procedure}
\vspace{-2mm}
\end{table}

%% file: tables/real-world.tex
\begin{table}[htbp]
\scriptsize
\caption{Top 5 cross-chain bridge in terms of asset affected by \bugname{} }

\begin{tabular}{l|ccccc}
\hline
Bridge               & \multicolumn{1}{l}{Hop.Exchanxx bridge} & \multicolumn{1}{l}{Terxx Bridge} & \multicolumn{1}{l}{Sifchxxx Bridge} & \multicolumn{1}{l}{RenBridxx} & \multicolumn{1}{l}{Ocuxx Bridge} \\ \hline
Vulnerability number & 4                                & 6                                & 1                                   & 3                             & 3                                \\
Affected asset       & 1445827                          & 28038.67                         & 16743.31                            & 12896.52                      & 8093.08                          \\ \hline
\end{tabular}

\label{table.bridge}
\vspace{-2mm}
\end{table}

%% file: 00_main.bbl

\begin{thebibliography}{44}


\ifx \showCODEN    \undefined \def \showCODEN     #1{\unskip}     \fi
\ifx \showDOI      \undefined \def \showDOI       #1{#1}\fi
\ifx \showISBNx    \undefined \def \showISBNx     #1{\unskip}     \fi
\ifx \showISBNxiii \undefined \def \showISBNxiii  #1{\unskip}     \fi
\ifx \showISSN     \undefined \def \showISSN      #1{\unskip}     \fi
\ifx \showLCCN     \undefined \def \showLCCN      #1{\unskip}     \fi
\ifx \shownote     \undefined \def \shownote      #1{#1}          \fi
\ifx \showarticletitle \undefined \def \showarticletitle #1{#1}   \fi
\ifx \showURL      \undefined \def \showURL       {\relax}        \fi
\providecommand\bibfield[2]{#2}
\providecommand\bibinfo[2]{#2}
\providecommand\natexlab[1]{#1}
\providecommand\showeprint[2][]{arXiv:#2}

\bibitem[Aafer et~al\mbox{.}(2018)]%
        {aafer2018AceDroid}
\bibfield{author}{\bibinfo{person}{Yousra Aafer}, \bibinfo{person}{Jianjun Huang}, \bibinfo{person}{Yi Sun}, \bibinfo{person}{Xiangyu Zhang}, \bibinfo{person}{Ninghui Li}, {and} \bibinfo{person}{Chen Tian}.} \bibinfo{year}{2018}\natexlab{}.
\newblock \showarticletitle{AceDroid: Normalizing Diverse Android Access Control Checks for Inconsistency Detection.}. In \bibinfo{booktitle}{\emph{25th Annual Network and Distributed System Security Symposium, NDSS 2018, San Diego, California, USA, February 18-21, 2018}}.
\newblock


\bibitem[Behnke(2022)]%
        {NOMADRootAttack}
\bibfield{author}{\bibinfo{person}{Rob Behnke}.} \bibinfo{year}{2022}\natexlab{}.
\newblock \bibinfo{title}{The nomad bridge hack: a deeper dive}.
\newblock \bibinfo{howpublished}{\url{https://www.halborn.com/blog/post/the-nomad-bridge-hack-a-deeper-dive}}.
\newblock
\newblock
\shownote{[Accessed 20-Sep-2023]}.


\bibitem[Binance(2022)]%
        {BNB}
\bibfield{author}{\bibinfo{person}{Binance}.} \bibinfo{year}{2022}\natexlab{}.
\newblock \bibinfo{title}{{BNB Chain}}.
\newblock \bibinfo{howpublished}{\url{https://www.bnbchain.org/}}.
\newblock
\newblock
\shownote{[Accessed 20-Sep-2023]}.


\bibitem[Bitcoin(2009)]%
        {Bitcoin}
\bibfield{author}{\bibinfo{person}{Bitcoin}.} \bibinfo{year}{2009}\natexlab{}.
\newblock \bibinfo{title}{Bitcoin}.
\newblock \bibinfo{howpublished}{\url{https://bitcoin.org/}}.
\newblock
\newblock
\shownote{[Accessed 20-Sep-2023]}.


\bibitem[Bose et~al\mbox{.}(2022)]%
        {bose2022SAILFISHa}
\bibfield{author}{\bibinfo{person}{Priyanka Bose}, \bibinfo{person}{Dipanjan Das}, \bibinfo{person}{Yanju Chen}, \bibinfo{person}{Yu Feng}, \bibinfo{person}{Christopher Kruegel}, {and} \bibinfo{person}{Giovanni Vigna}.} \bibinfo{year}{2022}\natexlab{}.
\newblock \showarticletitle{SAILFISH: Vetting Smart Contract State-Inconsistency Bugs in Seconds.}. In \bibinfo{booktitle}{\emph{43rd IEEE Symposium on Security and Privacy, SP 2022, San Francisco, CA, USA, May 22-26, 2022}}. \bibinfo{pages}{161--178}.
\newblock
\urldef\tempurl%
\url{https://doi.org/10.1109/SP46214.2022.9833721}
\showDOI{\tempurl}


\bibitem[Brent et~al\mbox{.}(2020)]%
        {brent2020Ethainter}
\bibfield{author}{\bibinfo{person}{Lexi Brent}, \bibinfo{person}{Neville Grech}, \bibinfo{person}{Sifis Lagouvardos}, \bibinfo{person}{Bernhard Scholz}, {and} \bibinfo{person}{Yannis Smaragdakis}.} \bibinfo{year}{2020}\natexlab{}.
\newblock \showarticletitle{Ethainter: A Smart Contract Security Analyzer for Composite Vulnerabilities}. In \bibinfo{booktitle}{\emph{Proceedings of the 41st ACM SIGPLAN Conference on Programming Language Design and Implementation}}. \bibinfo{publisher}{ACM}, \bibinfo{address}{London UK}, \bibinfo{pages}{454--469}.
\newblock
\showISBNx{978-1-4503-7613-6}
\urldef\tempurl%
\url{https://doi.org/10.1145/3385412.3385990}
\showDOI{\tempurl}


\bibitem[{CelerNetwork}(2022)]%
        {CelerNetworkDNSattack}
\bibfield{author}{\bibinfo{person}{{CelerNetwork}}.} \bibinfo{year}{2022}\natexlab{}.
\newblock \bibinfo{title}{A DNS cache poisoning attack on cBridge}.
\newblock \bibinfo{howpublished}{\url{https://twitter.com/CelerNetwork/status/1560123830844411904}}.
\newblock
\newblock
\shownote{[Accessed 20-Sep-2023]}.


\bibitem[Chainspot(2023)]%
        {chainspot}
\bibfield{author}{\bibinfo{person}{Chainspot}.} \bibinfo{year}{2023}\natexlab{}.
\newblock \bibinfo{title}{Chainspot}.
\newblock \bibinfo{howpublished}{\url{https://chainspot.io/}}.
\newblock
\newblock
\shownote{[Accessed 20-Sep-2023]}.


\bibitem[Choi et~al\mbox{.}(2021)]%
        {choi2021SMARTIAN}
\bibfield{author}{\bibinfo{person}{Jaeseung Choi}, \bibinfo{person}{Doyeon Kim}, \bibinfo{person}{Soomin Kim}, \bibinfo{person}{Gustavo Grieco}, \bibinfo{person}{Alex Groce}, {and} \bibinfo{person}{Sang~Kil Cha}.} \bibinfo{year}{2021}\natexlab{}.
\newblock \showarticletitle{SMARTIAN: Enhancing Smart Contract Fuzzing with Static and Dynamic Data-Flow Analyses}. In \bibinfo{booktitle}{\emph{2021 36th IEEE/ACM International Conference on Automated Software Engineering (ASE)}}. \bibinfo{publisher}{IEEE}, \bibinfo{address}{Melbourne, Australia}, \bibinfo{pages}{227--239}.
\newblock
\showISBNx{978-1-66540-337-5}
\urldef\tempurl%
\url{https://doi.org/10.1109/ASE51524.2021.9678888}
\showDOI{\tempurl}


\bibitem[Consensys(2017)]%
        {mythril}
\bibfield{author}{\bibinfo{person}{Consensys}.} \bibinfo{year}{2017}\natexlab{}.
\newblock \bibinfo{title}{Mythril}.
\newblock \bibinfo{howpublished}{\url{https://github.com/Consensys/mythril}}.
\newblock
\newblock
\shownote{[Accessed 20-Sep-2023]}.


\bibitem[Duan et~al\mbox{.}(2023)]%
        {duan2023Attacks}
\bibfield{author}{\bibinfo{person}{Li Duan}, \bibinfo{person}{Yangyang Sun}, \bibinfo{person}{Wei Ni}, \bibinfo{person}{Weiping Ding}, \bibinfo{person}{Jiqiang Liu}, {and} \bibinfo{person}{Wei Wang}.} \bibinfo{year}{2023}\natexlab{}.
\newblock \showarticletitle{Attacks Against Cross-Chain Systems and Defense Approaches: A Contemporary Survey}.
\newblock \bibinfo{journal}{\emph{IEEE/CAA Journal of Automatica Sinica}} \bibinfo{volume}{10}, \bibinfo{number}{8} (\bibinfo{date}{Aug.} \bibinfo{year}{2023}), \bibinfo{pages}{1647--1667}.
\newblock
\showISSN{2329-9266, 2329-9274}
\urldef\tempurl%
\url{https://doi.org/10.1109/JAS.2023.123642}
\showDOI{\tempurl}


\bibitem[{El-Rewini} et~al\mbox{.}(2022)]%
        {el-rewini2022Poirot}
\bibfield{author}{\bibinfo{person}{Zeinab {El-Rewini}}, \bibinfo{person}{Zhuo Zhang}, {and} \bibinfo{person}{Yousra Aafer}.} \bibinfo{year}{2022}\natexlab{}.
\newblock \showarticletitle{Poirot: Probabilistically Recommending Protections for the Android Framework.}. In \bibinfo{booktitle}{\emph{Proceedings of the 2022 ACM SIGSAC Conference on Computer and Communications Security, CCS 2022, Los Angeles, CA, USA, November 7-11, 2022}}. \bibinfo{pages}{937--950}.
\newblock
\urldef\tempurl%
\url{https://doi.org/10.1145/3548606.3560710}
\showDOI{\tempurl}


\bibitem[Ethereum(2015)]%
        {Ethereum}
\bibfield{author}{\bibinfo{person}{Ethereum}.} \bibinfo{year}{2015}\natexlab{}.
\newblock \bibinfo{title}{Ethereum}.
\newblock \bibinfo{howpublished}{\url{https://www.ethereum.org/}}.
\newblock
\newblock
\shownote{[Accessed 20-Sep-2023]}.


\bibitem[Feist et~al\mbox{.}(2019)]%
        {feist2019Slither}
\bibfield{author}{\bibinfo{person}{Josselin Feist}, \bibinfo{person}{Gustavo Grieco}, {and} \bibinfo{person}{Alex Groce}.} \bibinfo{year}{2019}\natexlab{}.
\newblock \showarticletitle{Slither: A Static Analysis Framework for Smart Contracts.}. In \bibinfo{booktitle}{\emph{Proceedings of the 2nd International Workshop on Emerging Trends in Software Engineering for Blockchain, WETSEB@ICSE 2019, Montreal, QC, Canada, May 27, 2019.}} \bibinfo{pages}{8--15}.
\newblock
\urldef\tempurl%
\url{https://doi.org/10.1109/WETSEB.2019.00008}
\showDOI{\tempurl}


\bibitem[Ghaleb et~al\mbox{.}(2022)]%
        {ghaleb2022ETainterb}
\bibfield{author}{\bibinfo{person}{Asem Ghaleb}, \bibinfo{person}{Julia Rubin}, {and} \bibinfo{person}{Karthik Pattabiraman}.} \bibinfo{year}{2022}\natexlab{}.
\newblock \showarticletitle{eTainter: Detecting Gas-Related Vulnerabilities in Smart Contracts}. In \bibinfo{booktitle}{\emph{Proceedings of the 31st ACM SIGSOFT International Symposium on Software Testing and Analysis}}. \bibinfo{publisher}{ACM}, \bibinfo{address}{Virtual South Korea}, \bibinfo{pages}{728--739}.
\newblock
\showISBNx{978-1-4503-9379-9}
\urldef\tempurl%
\url{https://doi.org/10.1145/3533767.3534378}
\showDOI{\tempurl}


\bibitem[Ghaleb et~al\mbox{.}(2023)]%
        {ghaleb2023AChecker}
\bibfield{author}{\bibinfo{person}{Asem Ghaleb}, \bibinfo{person}{Julia Rubin}, {and} \bibinfo{person}{Karthik Pattabiraman}.} \bibinfo{year}{2023}\natexlab{}.
\newblock \showarticletitle{AChecker: Statically Detecting Smart Contract Access Control Vulnerabilities}. In \bibinfo{booktitle}{\emph{2023 IEEE/ACM 45th International Conference on Software Engineering (ICSE)}}. \bibinfo{publisher}{IEEE}, \bibinfo{address}{Melbourne, Australia}, \bibinfo{pages}{945--956}.
\newblock
\showISBNx{978-1-66545-701-9}
\urldef\tempurl%
\url{https://doi.org/10.1109/ICSE48619.2023.00087}
\showDOI{\tempurl}


\bibitem[Grech et~al\mbox{.}(2018)]%
        {grech2018Madmax}
\bibfield{author}{\bibinfo{person}{Neville Grech}, \bibinfo{person}{Michael Kong}, \bibinfo{person}{Anton Jurisevic}, \bibinfo{person}{Lexi Brent}, \bibinfo{person}{Bernhard Scholz}, {and} \bibinfo{person}{Yannis Smaragdakis}.} \bibinfo{year}{2018}\natexlab{}.
\newblock \showarticletitle{Madmax: Surviving out-of-Gas Conditions in Ethereum Smart Contracts}.
\newblock \bibinfo{journal}{\emph{Proceedings of the ACM on Programming Languages}} \bibinfo{volume}{2}, \bibinfo{number}{OOPSLA} (\bibinfo{year}{2018}), \bibinfo{pages}{1--27}.
\newblock
\showISBNx{2475-1421}


\bibitem[Grieco et~al\mbox{.}(2020)]%
        {grieco2020Echidna}
\bibfield{author}{\bibinfo{person}{Gustavo Grieco}, \bibinfo{person}{Will Song}, \bibinfo{person}{Artur Cygan}, \bibinfo{person}{Josselin Feist}, {and} \bibinfo{person}{Alex Groce}.} \bibinfo{year}{2020}\natexlab{}.
\newblock \showarticletitle{Echidna: Effective, Usable, and Fast Fuzzing for Smart Contracts.}. In \bibinfo{booktitle}{\emph{ISSTA '20: 29th ACM SIGSOFT International Symposium on Software Testing and Analysis, Virtual Event, USA, July 18-22, 2020}}. \bibinfo{pages}{557--560}.
\newblock
\urldef\tempurl%
\url{https://doi.org/10.1145/3395363.3404366}
\showDOI{\tempurl}


\bibitem[Jiang et~al\mbox{.}(2018)]%
        {jiang2018ContractFuzzer}
\bibfield{author}{\bibinfo{person}{Bo Jiang}, \bibinfo{person}{Ye Liu}, {and} \bibinfo{person}{W.~K. Chan}.} \bibinfo{year}{2018}\natexlab{}.
\newblock \showarticletitle{ContractFuzzer: Fuzzing Smart Contracts for Vulnerability Detection}. In \bibinfo{booktitle}{\emph{Proceedings of the 33rd ACM/IEEE International Conference on Automated Software Engineering}}. \bibinfo{publisher}{ACM}, \bibinfo{address}{Montpellier France}, \bibinfo{pages}{259--269}.
\newblock
\showISBNx{978-1-4503-5937-5}
\urldef\tempurl%
\url{https://doi.org/10.1145/3238147.3238177}
\showDOI{\tempurl}


\bibitem[Kalra et~al\mbox{.}(2018)]%
        {kalra2018ZEUS}
\bibfield{author}{\bibinfo{person}{Sukrit Kalra}, \bibinfo{person}{Seep Goel}, \bibinfo{person}{Mohan Dhawan}, {and} \bibinfo{person}{Subodh Sharma}.} \bibinfo{year}{2018}\natexlab{}.
\newblock \showarticletitle{ZEUS: Analyzing Safety of Smart Contracts.}. In \bibinfo{booktitle}{\emph{25th Annual Network and Distributed System Security Symposium, NDSS 2018, San Diego, California, USA, February 18-21, 2018}}.
\newblock


\bibitem[Lee et~al\mbox{.}(2023)]%
        {lee2023SoK}
\bibfield{author}{\bibinfo{person}{Sung-Shine Lee}, \bibinfo{person}{Alexandr Murashkin}, \bibinfo{person}{Martin Derka}, {and} \bibinfo{person}{Jan Gorzny}.} \bibinfo{year}{2023}\natexlab{}.
\newblock \showarticletitle{SoK: Not Quite Water Under the Bridge: Review of Cross-Chain Bridge Hacks}. In \bibinfo{booktitle}{\emph{2023 IEEE International Conference on Blockchain and Cryptocurrency (ICBC)}}. \bibinfo{publisher}{IEEE}, \bibinfo{address}{Dubai, United Arab Emirates}, \bibinfo{pages}{1--14}.
\newblock
\showISBNx{9798350310191}
\urldef\tempurl%
\url{https://doi.org/10.1109/ICBC56567.2023.10174993}
\showDOI{\tempurl}


\bibitem[Liao et~al\mbox{.}(2023)]%
        {liao2023SmartState}
\bibfield{author}{\bibinfo{person}{Zeqin Liao}, \bibinfo{person}{Sicheng Hao}, \bibinfo{person}{Yuhong Nan}, {and} \bibinfo{person}{Zibin Zheng}.} \bibinfo{year}{2023}\natexlab{}.
\newblock \showarticletitle{SmartState: Detecting State-Reverting Vulnerabilities in Smart Contracts via Fine-Grained State-Dependency Analysis.}. In \bibinfo{booktitle}{\emph{Proceedings of the 32nd ACM SIGSOFT International Symposium on Software Testing and Analysis, ISSTA 2023, Seattle, WA, USA, July 17-21, 2023}}. \bibinfo{pages}{980--991}.
\newblock
\urldef\tempurl%
\url{https://doi.org/10.1145/3597926.3598111}
\showDOI{\tempurl}


\bibitem[Liao et~al\mbox{.}(2022)]%
        {liao2022SmartDagger}
\bibfield{author}{\bibinfo{person}{Zeqin Liao}, \bibinfo{person}{Zibin Zheng}, \bibinfo{person}{Xiao Chen}, {and} \bibinfo{person}{Yuhong Nan}.} \bibinfo{year}{2022}\natexlab{}.
\newblock \showarticletitle{SmartDagger: A Bytecode-Based Static Analysis Approach for Detecting Cross-Contract Vulnerability.}. In \bibinfo{booktitle}{\emph{ISSTA '22: 31st ACM SIGSOFT International Symposium on Software Testing and Analysis, Virtual Event, South Korea, July 18 - 22, 2022}}. \bibinfo{pages}{752--764}.
\newblock
\urldef\tempurl%
\url{https://doi.org/10.1145/3533767.3534222}
\showDOI{\tempurl}


\bibitem[Liu et~al\mbox{.}(2022)]%
        {liu2022Findinga}
\bibfield{author}{\bibinfo{person}{Ye Liu}, \bibinfo{person}{Yi Li}, \bibinfo{person}{Shang-Wei Lin}, {and} \bibinfo{person}{Cyrille Artho}.} \bibinfo{year}{2022}\natexlab{}.
\newblock \showarticletitle{Finding Permission Bugs in Smart Contracts with Role Mining.}. In \bibinfo{booktitle}{\emph{ISSTA '22: 31st ACM SIGSOFT International Symposium on Software Testing and Analysis, Virtual Event, South Korea, July 18 - 22, 2022}}. \bibinfo{pages}{716--727}.
\newblock
\urldef\tempurl%
\url{https://doi.org/10.1145/3533767.3534372}
\showDOI{\tempurl}


\bibitem[Liu et~al\mbox{.}(2021)]%
        {liu2021Smart}
\bibfield{author}{\bibinfo{person}{Zhenguang Liu}, \bibinfo{person}{Peng Qian}, \bibinfo{person}{Xiang Wang}, \bibinfo{person}{Lei Zhu}, \bibinfo{person}{Qinming He}, {and} \bibinfo{person}{Shouling Ji}.} \bibinfo{year}{2021}\natexlab{}.
\newblock \showarticletitle{Smart Contract Vulnerability Detection: From Pure Neural Network to Interpretable Graph Feature and Expert Pattern Fusion}. In \bibinfo{booktitle}{\emph{Proceedings of the Thirtieth International Joint Conference on Artificial Intelligence}}. \bibinfo{publisher}{International Joint Conferences on Artificial Intelligence Organization}, \bibinfo{address}{Montreal, Canada}, \bibinfo{pages}{2751--2759}.
\newblock
\showISBNx{978-0-9992411-9-6}
\urldef\tempurl%
\url{https://doi.org/10.24963/ijcai.2021/379}
\showDOI{\tempurl}


\bibitem[Luu et~al\mbox{.}(2016)]%
        {luu2016Making}
\bibfield{author}{\bibinfo{person}{Loi Luu}, \bibinfo{person}{Duc-Hiep Chu}, \bibinfo{person}{Hrishi Olickel}, \bibinfo{person}{Prateek Saxena}, {and} \bibinfo{person}{Aquinas Hobor}.} \bibinfo{year}{2016}\natexlab{}.
\newblock \showarticletitle{Making Smart Contracts Smarter}. In \bibinfo{booktitle}{\emph{Proceedings of the 2016 ACM SIGSAC Conference on Computer and Communications Security}}. \bibinfo{publisher}{ACM}, \bibinfo{address}{Vienna Austria}, \bibinfo{pages}{254--269}.
\newblock
\showISBNx{978-1-4503-4139-4}
\urldef\tempurl%
\url{https://doi.org/10.1145/2976749.2978309}
\showDOI{\tempurl}


\bibitem[Network(2020)]%
        {PolyNetwork}
\bibfield{author}{\bibinfo{person}{Poly Network}.} \bibinfo{year}{2020}\natexlab{}.
\newblock \bibinfo{title}{{PolyNetwork}}.
\newblock \bibinfo{howpublished}{\url{https://www.poly.network/}}.
\newblock
\newblock
\shownote{[Accessed 20-Sep-2023]}.


\bibitem[Nguyen et~al\mbox{.}(2020)]%
        {nguyen2020SFuzz}
\bibfield{author}{\bibinfo{person}{Tai~D. Nguyen}, \bibinfo{person}{Long~H. Pham}, \bibinfo{person}{Jun Sun}, \bibinfo{person}{Yun Lin}, {and} \bibinfo{person}{Quang~Tran Minh}.} \bibinfo{year}{2020}\natexlab{}.
\newblock \showarticletitle{sFuzz: An Efficient Adaptive Fuzzer for Solidity Smart Contracts}. In \bibinfo{booktitle}{\emph{Proceedings of the ACM/IEEE 42nd International Conference on Software Engineering}}. \bibinfo{publisher}{ACM}, \bibinfo{address}{Seoul South Korea}, \bibinfo{pages}{778--788}.
\newblock
\showISBNx{978-1-4503-7121-6}
\urldef\tempurl%
\url{https://doi.org/10.1145/3377811.3380334}
\showDOI{\tempurl}


\bibitem[Polygon(2023)]%
        {polygonbridge}
\bibfield{author}{\bibinfo{person}{Polygon}.} \bibinfo{year}{2023}\natexlab{}.
\newblock \bibinfo{title}{Polygon Bridge}.
\newblock \bibinfo{howpublished}{\url{https://chainspot.io/bridge/polygon-bridge}}.
\newblock
\newblock
\shownote{[Accessed 20-Sep-2023]}.


\bibitem[{Rubic}(2022)]%
        {RubicPrivateleaking}
\bibfield{author}{\bibinfo{person}{{Rubic}}.} \bibinfo{year}{2022}\natexlab{}.
\newblock \bibinfo{title}{Rubic admin wallet addresses was compromised}.
\newblock \bibinfo{howpublished}{\url{https://twitter.com/CryptoRubic/status/1587704548688367619}}.
\newblock
\newblock
\shownote{[Accessed 20-Sep-2023]}.


\bibitem[{Sam Cooling}(2021)]%
        {ChainSwap}
\bibfield{author}{\bibinfo{person}{{Sam Cooling}}.} \bibinfo{year}{2021}\natexlab{}.
\newblock \bibinfo{title}{ChainSwap hackers steal \$8m and crash token prices}.
\newblock \bibinfo{howpublished}{\url{https://finance.yahoo.com/news/chainswap-hackers-steal-8m-crash-121056965.html}}.
\newblock
\newblock
\shownote{[Accessed 20-Sep-2023]}.


\bibitem[{Sebastian Sinclair}(2021)]%
        {Thorchain}
\bibfield{author}{\bibinfo{person}{{Sebastian Sinclair}}.} \bibinfo{year}{2021}\natexlab{}.
\newblock \bibinfo{title}{Blockchain Protocol Thorchain Suffers \$8M Hack}.
\newblock \bibinfo{howpublished}{\url{https://www.coindesk.com/markets/2021/07/23/blockchain-protocol-thorchain-suffers-8m-hack/}}.
\newblock
\newblock
\shownote{[Accessed 20-Sep-2023]}.


\bibitem[Shao et~al\mbox{.}(2016)]%
        {shao2016Kratos}
\bibfield{author}{\bibinfo{person}{Yuru Shao}, \bibinfo{person}{Qi~Alfred Chen}, \bibinfo{person}{Zhuoqing~Morley Mao}, \bibinfo{person}{Jason Ott}, {and} \bibinfo{person}{Zhiyun Qian}.} \bibinfo{year}{2016}\natexlab{}.
\newblock \showarticletitle{Kratos: Discovering Inconsistent Security Policy Enforcement in the Android Framework.}. In \bibinfo{booktitle}{\emph{23rd Annual Network and Distributed System Security Symposium, NDSS 2016, San Diego, California, USA, February 21-24, 2016}}.
\newblock


\bibitem[Shou et~al\mbox{.}(2023)]%
        {shou2023ItyFuzz}
\bibfield{author}{\bibinfo{person}{Chaofan Shou}, \bibinfo{person}{Shangyin Tan}, {and} \bibinfo{person}{Koushik Sen}.} \bibinfo{year}{2023}\natexlab{}.
\newblock \showarticletitle{ItyFuzz: Snapshot-Based Fuzzer for Smart Contract.}. In \bibinfo{booktitle}{\emph{Proceedings of the 32nd ACM SIGSOFT International Symposium on Software Testing and Analysis, ISSTA 2023, Seattle, WA, USA, July 17-21, 2023}}. \bibinfo{pages}{322--333}.
\newblock
\urldef\tempurl%
\url{https://doi.org/10.1145/3597926.3598059}
\showDOI{\tempurl}


\bibitem[Tsankov et~al\mbox{.}(2018)]%
        {tsankov2018Securify}
\bibfield{author}{\bibinfo{person}{Petar Tsankov}, \bibinfo{person}{Andrei~Marian Dan}, \bibinfo{person}{Dana {Drachsler-Cohen}}, \bibinfo{person}{Arthur Gervais}, \bibinfo{person}{Florian B{\"u}nzli}, {and} \bibinfo{person}{Martin~T. Vechev}.} \bibinfo{year}{2018}\natexlab{}.
\newblock \showarticletitle{Securify: Practical Security Analysis of Smart Contracts.}. In \bibinfo{booktitle}{\emph{Proceedings of the 2018 ACM SIGSAC Conference on Computer and Communications Security, CCS 2018, Toronto, ON, Canada, October 15-19, 2018}}. \bibinfo{pages}{67--82}.
\newblock
\urldef\tempurl%
\url{https://doi.org/10.1145/3243734.3243780}
\showDOI{\tempurl}


\bibitem[Vishwakarma et~al\mbox{.}(2023)]%
        {vishwakarma2023CrossLedger}
\bibfield{author}{\bibinfo{person}{Lokendra Vishwakarma}, \bibinfo{person}{Amritesh Kumar}, {and} \bibinfo{person}{Debasis Das}.} \bibinfo{year}{2023}\natexlab{}.
\newblock \showarticletitle{CrossLedger: A Pioneer Cross-Chain Asset Transfer Protocol}. In \bibinfo{booktitle}{\emph{2023 IEEE/ACM 23rd International Symposium on Cluster, Cloud and Internet Computing (CCGrid)}}. \bibinfo{publisher}{IEEE}, \bibinfo{address}{Bangalore, India}, \bibinfo{pages}{568--578}.
\newblock
\showISBNx{9798350301199}
\urldef\tempurl%
\url{https://doi.org/10.1109/CCGrid57682.2023.00059}
\showDOI{\tempurl}


\bibitem[Wikipedia(2023)]%
        {NFT}
\bibfield{author}{\bibinfo{person}{Wikipedia}.} \bibinfo{year}{2023}\natexlab{}.
\newblock \bibinfo{title}{Non fungible token}.
\newblock \bibinfo{howpublished}{\url{https://en.wikipedia.org/wiki/Non-fungible_token}}.
\newblock
\newblock
\shownote{[Accessed 20-Sep-2023]}.


\bibitem[{Wikipedia contributors}(2022)]%
        {PolyNetworkexploit}
\bibfield{author}{\bibinfo{person}{{Wikipedia contributors}}.} \bibinfo{year}{2022}\natexlab{}.
\newblock \bibinfo{title}{Poly Network exploit}.
\newblock \bibinfo{howpublished}{\url{https://en.wikipedia.org/wiki/Poly_Network_exploit}}.
\newblock
\newblock
\shownote{[Accessed 20-Sep-2023]}.


\bibitem[W{\"u}stholz and Christakis(2020)]%
        {wustholz2020Harvey}
\bibfield{author}{\bibinfo{person}{Valentin W{\"u}stholz} {and} \bibinfo{person}{Maria Christakis}.} \bibinfo{year}{2020}\natexlab{}.
\newblock \showarticletitle{Harvey: A Greybox Fuzzer for Smart Contracts.}. In \bibinfo{booktitle}{\emph{ESEC/FSE '20: 28th ACM Joint European Software Engineering Conference and Symposium on the Foundations of Software Engineering, Virtual Event, USA, November 8-13, 2020}}. \bibinfo{pages}{1398--1409}.
\newblock
\urldef\tempurl%
\url{https://doi.org/10.1145/3368089.3417064}
\showDOI{\tempurl}


\bibitem[Xie et~al\mbox{.}(2022)]%
        {xie2022ZkBridge}
\bibfield{author}{\bibinfo{person}{Tiancheng Xie}, \bibinfo{person}{Jiaheng Zhang}, \bibinfo{person}{Zerui Cheng}, \bibinfo{person}{Fan Zhang}, \bibinfo{person}{Yupeng Zhang}, \bibinfo{person}{Yongzheng Jia}, \bibinfo{person}{Dan Boneh}, {and} \bibinfo{person}{Dawn Song}.} \bibinfo{year}{2022}\natexlab{}.
\newblock \showarticletitle{zkBridge: Trustless Cross-Chain Bridges Made Practical}. In \bibinfo{booktitle}{\emph{Proceedings of the 2022 ACM SIGSAC Conference on Computer and Communications Security}}. \bibinfo{publisher}{ACM}, \bibinfo{address}{Los Angeles CA USA}, \bibinfo{pages}{3003--3017}.
\newblock
\showISBNx{978-1-4503-9450-5}
\urldef\tempurl%
\url{https://doi.org/10.1145/3548606.3560652}
\showDOI{\tempurl}


\bibitem[Xue et~al\mbox{.}(2020)]%
        {xue2020Crosscontract}
\bibfield{author}{\bibinfo{person}{Yinxing Xue}, \bibinfo{person}{Mingliang Ma}, \bibinfo{person}{Yun Lin}, \bibinfo{person}{Yulei Sui}, \bibinfo{person}{Jiaming Ye}, {and} \bibinfo{person}{Tianyong Peng}.} \bibinfo{year}{2020}\natexlab{}.
\newblock \showarticletitle{Cross-Contract Static Analysis for Detecting Practical Reentrancy Vulnerabilities in Smart Contracts}. In \bibinfo{booktitle}{\emph{Proceedings of the 35th IEEE/ACM International Conference on Automated Software Engineering}}. \bibinfo{publisher}{ACM}, \bibinfo{address}{Virtual Event Australia}, \bibinfo{pages}{1029--1040}.
\newblock
\showISBNx{978-1-4503-6768-4}
\urldef\tempurl%
\url{https://doi.org/10.1145/3324884.3416553}
\showDOI{\tempurl}


\bibitem[Zhang et~al\mbox{.}(2022)]%
        {zhang2022Xscope}
\bibfield{author}{\bibinfo{person}{Jiashuo Zhang}, \bibinfo{person}{Jianbo Gao}, \bibinfo{person}{Yue Li}, \bibinfo{person}{Ziming Chen}, \bibinfo{person}{Zhi Guan}, {and} \bibinfo{person}{Zhong Chen}.} \bibinfo{year}{2022}\natexlab{}.
\newblock \showarticletitle{Xscope: Hunting for Cross-Chain Bridge Attacks}. In \bibinfo{booktitle}{\emph{Proceedings of the 37th IEEE/ACM International Conference on Automated Software Engineering}}. \bibinfo{publisher}{ACM}, \bibinfo{address}{Rochester MI USA}, \bibinfo{pages}{1--4}.
\newblock
\showISBNx{978-1-4503-9475-8}
\urldef\tempurl%
\url{https://doi.org/10.1145/3551349.3559520}
\showDOI{\tempurl}


\bibitem[Zheng et~al\mbox{.}(2023)]%
        {zheng2023survey}
\bibfield{author}{\bibinfo{person}{Zibin Zheng}, \bibinfo{person}{Kaiwen Ning}, \bibinfo{person}{Yanlin Wang}, \bibinfo{person}{Jingwen Zhang}, \bibinfo{person}{Dewu Zheng}, \bibinfo{person}{Mingxi Ye}, {and} \bibinfo{person}{Jiachi Chen}.} \bibinfo{year}{2023}\natexlab{}.
\newblock \showarticletitle{A Survey of Large Language Models for Code: Evolution, Benchmarking, and Future Trends}.
\newblock \bibinfo{journal}{\emph{CoRR}}  \bibinfo{volume}{abs/2311.10372}.
\newblock
\urldef\tempurl%
\url{https://doi.org/10.48550/ARXIV.2311.10372}
\showDOI{\tempurl}


\bibitem[Zheng et~al\mbox{.}(2020)]%
        {zheng_overview_2020}
\bibfield{author}{\bibinfo{person}{Zibin Zheng}, \bibinfo{person}{Shaoan Xie}, \bibinfo{person}{Hong{-}Ning Dai}, \bibinfo{person}{Weili Chen}, \bibinfo{person}{Xiangping Chen}, \bibinfo{person}{Jian Weng}, {and} \bibinfo{person}{Muhammad Imran}.} \bibinfo{year}{2020}\natexlab{}.
\newblock \showarticletitle{An overview on smart contracts: Challenges, advances and platforms}.
\newblock \bibinfo{journal}{\emph{Future Gener. Comput. Syst.}}  \bibinfo{volume}{105} (\bibinfo{year}{2020}), \bibinfo{pages}{475--491}.
\newblock
\urldef\tempurl%
\url{https://doi.org/10.1016/J.FUTURE.2019.12.019}
\showDOI{\tempurl}


\end{thebibliography}
